\documentclass[11pt]{article}

\usepackage[preprint]{acl}

\usepackage{times}
\usepackage{latexsym}

\usepackage[T1]{fontenc}
\usepackage[utf8]{inputenc}

\usepackage{microtype}
\usepackage{inconsolata}

\usepackage{graphicx}

\usepackage{mathtools}
\usepackage{amssymb}
\usepackage{amsthm}

\usepackage{siunitx}

\usepackage{booktabs}
\usepackage{multirow}
\usepackage{tabularx}
\usepackage{array}
\usepackage{xcolor}
\usepackage{colortbl}
\usepackage{placeins}

\usepackage{float}
\usepackage{afterpage}
\usepackage{textcomp}
\usepackage{gensymb}
\usepackage{enumitem}
\usepackage{subcaption}

\usepackage{algorithm}
\usepackage[noend]{algpseudocode}

\newtheoremstyle{boldstyle}
  {\topsep}
  {\topsep}
  {\itshape}
  {}
  {\bfseries}
  {.}
  {.5em}
  {}

\theoremstyle{boldstyle}
\newtheorem{theorem}{Theorem}
\newtheorem{lemma}[theorem]{Lemma}
\newtheorem{proposition}[theorem]{Proposition}
\newtheorem{definition}[theorem]{Definition}
\newtheorem{remark}[theorem]{Remark}
\newtheorem{assumption}{Assumption}
\newtheorem{corollary}{Corollary}

\newcommand{\SPD}{\mathrm{SPD}}
\newcommand{\Tail}{\mathrm{Tail}}
\newcommand{\Dgeom}{\Delta_{\mathrm{geom}}}
\newcommand{\Dtail}{\Delta_{\mathrm{tail}}}

\newcolumntype{L}[1]{>{\raggedright\arraybackslash}m{#1}}
\newcolumntype{C}[1]{>{\centering\arraybackslash}m{#1}}
\newcolumntype{R}[1]{>{\raggedleft\arraybackslash}m{#1}}

\algrenewcommand\algorithmicrequire{\textbf{Require:}}
\algrenewcommand\algorithmicensure{\textbf{Ensure:}}

\hypersetup{
  colorlinks=true,
  linkcolor=black,
  citecolor=blue,
  urlcolor=blue,
  pdfborder={0 0 0}
}

\title{Bridging Distance and Spectral Positional Encodings via Anchor-Based Diffusion Geometry Approximation}

\author{
  Zimo Yan \and Zheng Xie\thanks{Corresponding author: Zheng Xie (xiezheng81@nudt.edu.cn).} \and
  Runfan Duan \and Chang Liu \and Wumei Du \\
  National University of Defense Technology, Changsha, China \\
  \texttt{\{yanzimo20,xiezheng81,duanrunfan24,liuchang\_,wumeidu\}@nudt.edu.cn}
}

\begin{document}
\maketitle
\begin{abstract}
Molecular graph learning benefits from positional signals that capture both local neighborhoods and global topology. Two widely used families are spectral encodings derived from Laplacian or diffusion operators and anchor-based distance encodings built from shortest-path information, yet their precise relationship is poorly understood. We interpret distance encodings as a low-rank surrogate of diffusion geometry and derive an explicit trilateration map that reconstructs truncated diffusion coordinates from transformed anchor distances and anchor spectral positions, with pointwise and Frobenius-gap guarantees on random regular graphs. On DrugBank molecular graphs using a shared GNP-based DDI prediction backbone, a distance-driven Nystr\"om scheme closely recovers diffusion geometry, and both Laplacian and distance encodings substantially outperform a no-encoding baseline.
\end{abstract}

\section{Introduction}

Molecular graph modeling tasks--from de novo graph generation to molecular reasoning--require representations that capture both local chemical neighborhoods and global topology \citeyearpar{bronstein2017geometric,zhou2020graph,wu2020comprehensive}. A recurring challenge is how to inject structural positional information so that long-range dependencies and global organization are accessible to the model. In this work we use Graph Neural Processes (GNPs) as a flexible probabilistic backbone \citeyearpar{yan2025metamolgenneuralgraphmotif}.

Many positional/structural signals have been explored, including random-walk/propagation statistics, structural-role descriptors, and relative-position biases in graph transformers \citeyearpar{rampasek2022recipe,eliasof2023graph,ying2021transformers}. We focus on two particularly common families. Spectral/diffusion encodings (e.g., Laplacian eigenmaps, heat-kernel and diffusion-map features) provide coordinates aligned with diffusion geometry but often rely on costly eigendecomposition or careful approximation \citeyearpar{vonluxburg2007tutorial,dwivedi2023benchmarking,maskey2022generalized}. Distance/anchor encodings represent each node by its shortest-path distances to a small anchor set (optionally transformed), offering a simple and scalable alternative without explicit spectral computation \citeyearpar{yan2025resolvingnodeidentifiabilitygraph,li2009distance}.

Despite their widespread use, an explicit algebraic account of when anchor-based distance features can approximate diffusion geometry is still missing, as are error measures that are meaningful both for recovered coordinates and for induced distance/kernel matrices. This gap matters in molecular settings where diffusion geometry is a natural inductive bias and computational budgets are limited, raising the practical question of when distance encodings can serve as principled substitutes for spectral encodings in tasks such as drug-drug interaction (DDI) prediction \citeyearpar{zitnik2018modeling,ryu2018deep}.

\textbf{Motivation.}
We ask: \textit{Can we construct an explicit algebraic map from anchor-based shortest-path encodings to truncated diffusion (spectral) coordinates, with provable pointwise and matrix-level error guarantees?} Such a link would connect discrete distance primitives to diffusion geometry and clarify the accuracy-efficiency trade-off behind replacing spectral computation with distance features.

We address this question through theory and experiments. On the theory side, we derive an explicit trilateration operator that reconstructs truncated diffusion coordinates from transformed anchor distances and anchor spectral positions, and we establish pointwise and Frobenius-gap guarantees under a random regular graph model with a local monotone distance linkage. On the empirical side, we evaluate diffusion-geometry recovery on DrugBank using a distance-driven Nystr\"om scheme and assess downstream DDI prediction on DrugBank and ChCh-Miner under a shared GNP backbone \citeyearpar{lin2020kgnn,deng2020multimodal,ma2023dual}. Overall, both Laplacian and distance encodings improve over NoPE, with Laplacian encodings giving the most consistent gains.

Our main contributions are:
\begin{enumerate}
   \item We provide a spectral--algebraic bridge by formalizing anchor-based distance encodings as a low-rank surrogate for diffusion geometry and developing an explicit trilateration operator with pointwise and Frobenius-gap guarantees under random regular graph assumptions.
   \item We show that distance-based Nystr\"om approximations closely recover diffusion geometry on DrugBank molecular graphs, yielding accurate kernel and embedding approximations with a moderate number of anchors.
   \item Using a common GNP-based architecture for DDI prediction, we compare NoPE, DE, and LapPE, and conduct ablations on distance transforms and anchor counts on DrugBank and ChCh-Miner.
\end{enumerate}

\section{Literature Review}
\label{SE2}

This section reviews structural and positional information for graph learning, with an emphasis on molecular graphs. We discuss why purely local computation can be structurally limiting, summarize spectral/diffusion and anchor-based distance encodings as two common positional encoding families, and review drug-drug interaction (DDI) models to identify the gap addressed by our spectral-algebraic bridge between distance features and diffusion geometry.

\subsection{Graph structure learning and locality limitations}

Many graph learning methods adopt locality-based computation, most prominently message passing: node representations are iteratively updated by aggregating and transforming features from adjacent nodes, forming the backbone of many GNN architectures and related graph encoders \citeyearpar{yan2025metamolgenneuralgraphmotif, gilmer2017neural,kipf2016semi}. Such designs have shown strong empirical performance across a wide range of applications, including molecular property prediction.

Despite this success, locality imposes structural limitations. The expressive power of standard message passing is upper bounded by the Weisfeiler-Leman test, so non-isomorphic graphs and structurally distinct nodes can remain indistinguishable \citeyearpar{morris2019weisfeiler,xu2018powerful}. In addition, long-range dependencies can be difficult to capture due to over-smoothing and over-squashing effects \citeyearpar{alon2021on,topping2021understanding}. These issues motivate enriching node features with explicit structural or positional signals beyond local neighborhood aggregation.

\subsection{Spectral positional encodings}

Beyond spectral and distance-based designs, prior work also studies random-walk/propagation statistics, structural roles, transformer relative-position biases, and substructure-count descriptors \citeyearpar{rampasek2022recipe,eliasof2023graph,ying2021transformers}. We focus on spectral/diffusion and anchor-distance encodings because they provide a natural geometric lens and are the two families we explicitly connect.

Spectral positional encodings derive coordinates from Laplacian or diffusion-operator eigenvectors (e.g., Laplacian eigenmaps and diffusion maps) \citeyearpar{belkin2003laplacian,coifman2005geometric,coifman2006diffusion}, and are widely used as fixed Laplacian positional features to inject global topology and break symmetries \citeyearpar{dwivedi2023benchmarking}. Variants adapt or approximate these features via learned reweighting/combinations or propagation-based approximations \citeyearpar{maskey2022generalized,eliasof2023graph}, but spectral methods can still be costly on large or evolving graphs due to reliance on global modes.

\subsection{Distance based positional encodings}

Distance-based positional encodings represent each node by its shortest-path distances (or simple transforms) to a small set of anchor nodes \citeyearpar{li2009distance,you2019pgnn}. By relying on shortest-path computations rather than eigen-decomposition, these features are typically easy to implement and scale well with standard graph routines.

Such encodings have been used to enhance expressivity beyond Weisfeiler-Leman \citeyearpar{li2009distance} and to build structure-aware graph transformers through shortest-path distances and related structural priors \citeyearpar{ying2021transformers}. However, their connection to spectral/diffusion geometry is often only discussed qualitatively, leaving open when a finite anchor set can approximate diffusion geometry and how to quantify approximation error at both coordinate and kernel/distance levels.

\subsection{Graph based models for drug-drug interaction prediction}

Graph-based approaches have become central to drug-drug interaction (DDI) prediction. Early deep learning methods combined molecular fingerprints or sequence representations with standard neural architectures \citeyearpar{ryu2018deep}, while later work leveraged biomedical knowledge graphs and applied GNNs or knowledge-graph neural networks to model interactions in heterogeneous networks \citeyearpar{zitnik2018modeling,lin2020kgnn}. Multimodal frameworks further enriched drug representations by integrating molecular graphs with targets, pathways, and text \citeyearpar{deng2020multimodal}.

More recent architectures explicitly exploit molecular structure via dual-graph designs and co-attention mechanisms to capture substructure-level interactions between drug pairs \citeyearpar{yan2025multiscalegraphneuralprocess,feng2020interpreter}. These results consistently highlight the importance of structural information for accurate DDI prediction; nevertheless, most models adopt a fixed positional/structural feature design and emphasize architectural or data-integration choices, with comparatively limited work that systematically relates distance-based encodings to spectral diffusion geometry under a unified theoretical lens.

\subsection{Summary and remaining gap}

\begin{figure}[t]
  \centering
  \includegraphics[width=\columnwidth]{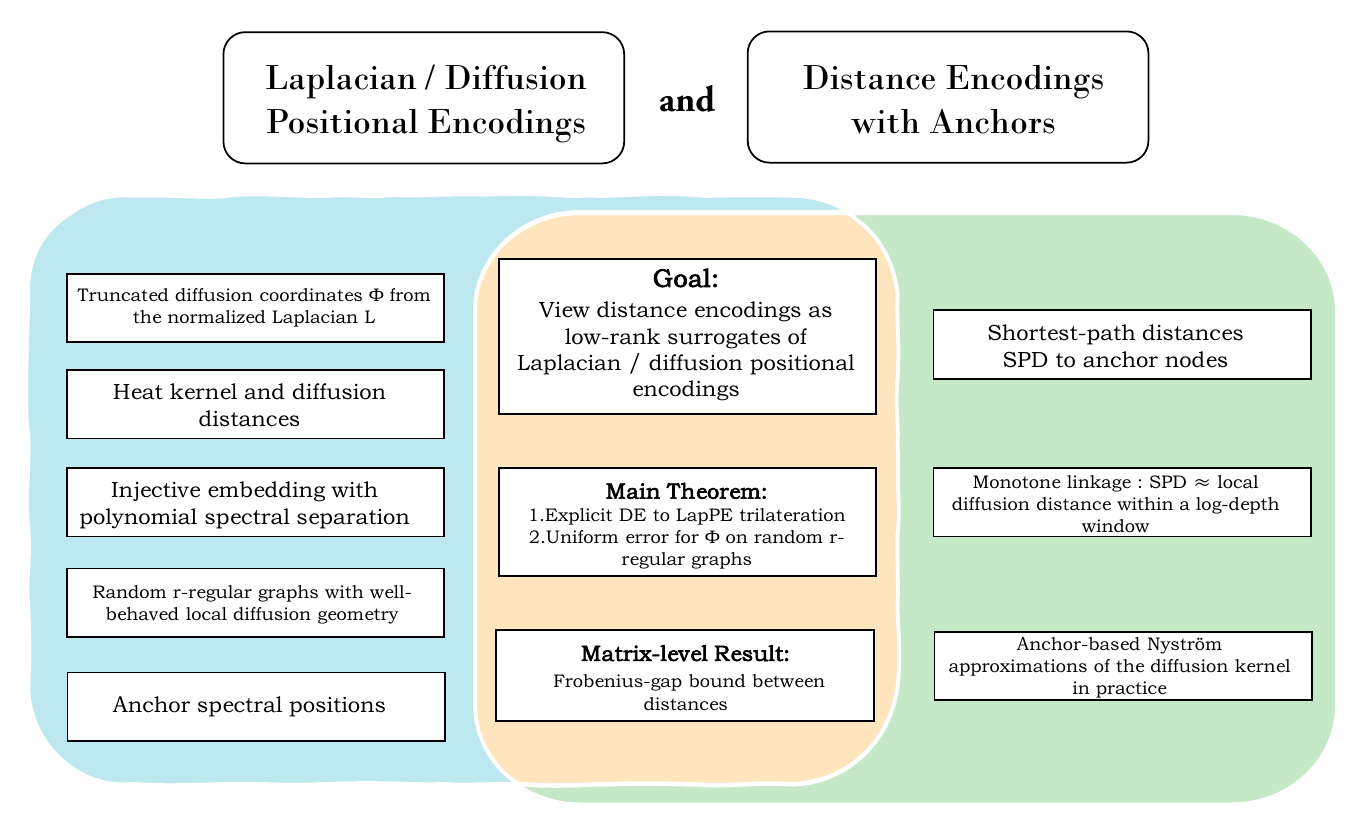}
  \caption{Overview of spectral/diffusion and anchor-distance positional encodings, and the algebraic bridge developed in this work.}
  \label{fig:theory_bridge}
\end{figure}

\begin{figure*}[t]
  \centering
  \includegraphics[width=0.85\textwidth]{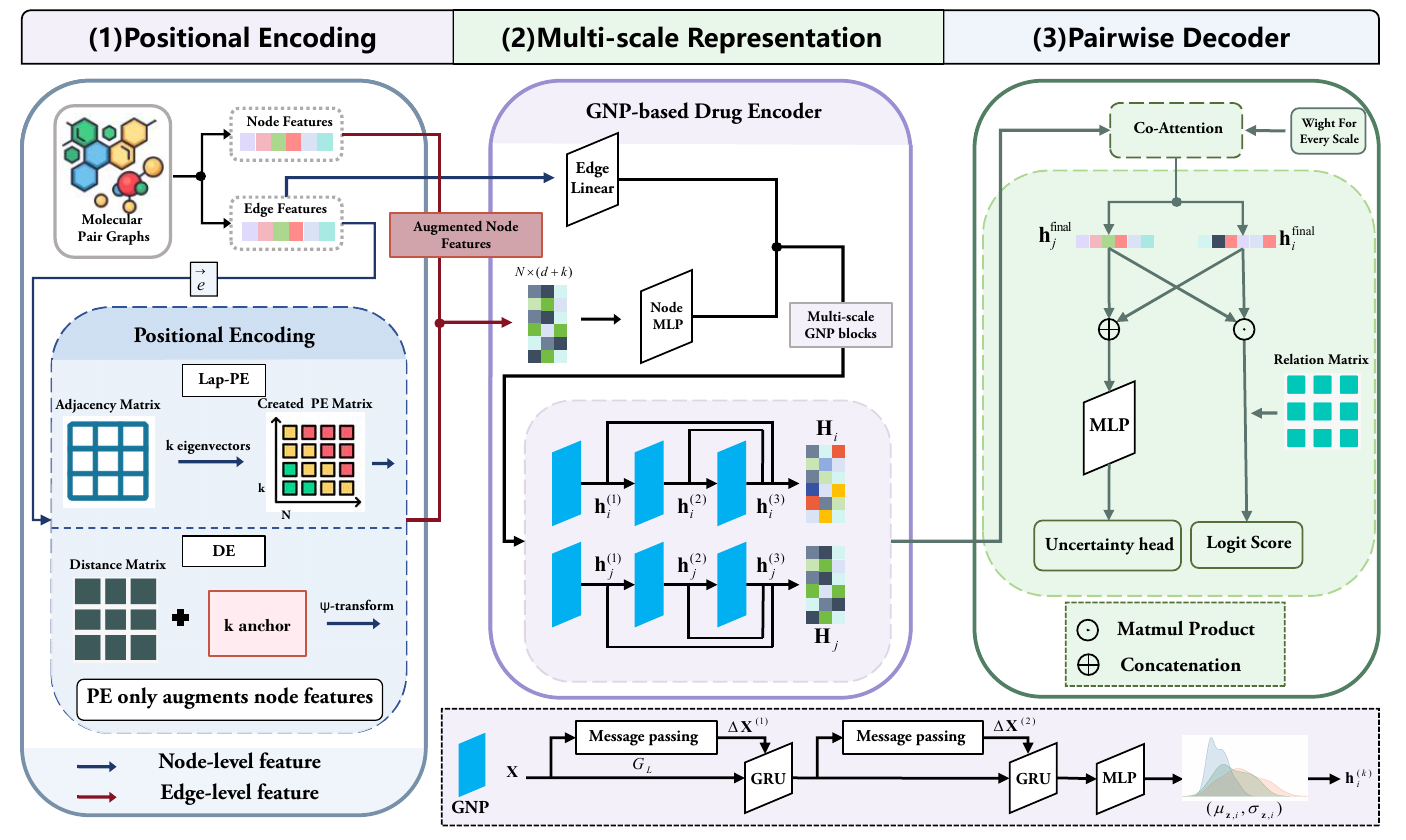}
  \caption{Proposed DDI framework: positional encodings augment molecular graphs, which are encoded by a shared multi-scale GNP and decoded via co-attention and relation-aware scoring.}
  \label{fig:model_architecture}
\end{figure*}

Spectral encodings yield geometrically meaningful but globally costly coordinates, whereas distance-based encodings are scalable and simple yet are typically justified empirically and lack a unified algebraic account of when they serve as low-rank surrogates for diffusion geometry. We close this gap by providing reconstruction operators and error bounds linking distance and spectral encodings and evaluating the resulting accuracy-efficiency trade-off for DDI prediction in a unified GNP-based framework (Figures~\ref{fig:theory_bridge} and~\ref{fig:model_architecture}).

\section{Preliminaries}
\label{SE3}
We collect notation and basic geometric objects used in the analysis, and formalize the two positional-encoding families, anchor-based shortest-path distance encodings and truncated Laplacian spectral coordinates, together with the algebraic comparison problem studied in Section~\ref{SE5}.

\subsection{Notations}
Let $G=(V,E)$ be a finite, connected, undirected graph with $|V|=n$. For $u,v\in V$, let $\SPD(u,v)$ denote the shortest-path distance. For $u\in V$ and $R\in\mathbb{N}$, define the radius-$R$ ball
\begin{equation}
 B_G(u,R)=\{\,v\in V:\SPD(u,v)\le R\,\}.
\end{equation}
Let $A\in\{0,1\}^{n\times n}$ be the adjacency matrix and $D$ the diagonal degree matrix. In the theoretical part we focus on $r$-regular graphs, so $D=rI$ and we use
\begin{equation}
 L := I-\frac{1}{r}A.
\end{equation}

Let $\{(\lambda_j,\varphi_j)\}_{j=1}^n$ be the eigenpairs of $L$, with $\{\varphi_j\}$ orthonormal in $\ell^2(V)$ and
$0=\lambda_1\le\lambda_2\le\cdots\le\lambda_n$. For $t>0$, define the heat semigroup
\begin{equation}
 K_t = e^{-tL},
\end{equation}
with entries $k_t(u,v)=(K_t)_{uv}$. The diffusion distance at time $t$ is equivalently given by
\begin{equation}
\begin{aligned}
  d_t(u,v)^2
 &= \sum_{j=1}^n e^{-2t\lambda_j}\bigl(\varphi_j(u)-\varphi_j(v)\bigr)^2\\
 &= k_{2t}(u,u)+k_{2t}(v,v)-2k_{2t}(u,v).   
\end{aligned}
\end{equation}

For an integer $m\ge 1$, define the truncated diffusion-map embedding
\begin{equation}
 \Phi_t^{(m)}(v)
 = \bigl(e^{-t\lambda_{j+1}}\varphi_{j+1}(v)\bigr)_{j=1}^m
 \in\mathbb{R}^m,
\end{equation}
and the corresponding truncated diffusion distance
\begin{equation}
 d_t^{(m)}(u,v)=\|\Phi_t^{(m)}(u)-\Phi_t^{(m)}(v)\|_2.
\end{equation}
Given anchors $a_1,\dots,a_{m+1}\in V$, write
\begin{equation}
 p_i=\Phi_t^{(m)}(a_i)\in\mathbb{R}^m,\quad i=1,\dots,m+1.
\end{equation}

\subsection{Problem Statement}
Fix a graph $G=(V,E)$ and an anchor set $S=\{a_1,\dots,a_{m+1}\}\subset V$. For any node $v\in V$, define the (transformed) distance encoding
\begin{equation}
\resizebox{\columnwidth}{!}{$
 \tilde{\zeta}(v\mid S)
 = \bigl(\psi(\SPD(a_1,v)),\dots,\psi(\SPD(a_{m+1},v))\bigr)^\top
 \in\mathbb{R}^{m+1},
 $}
\end{equation}
where $\psi:\mathbb{R}_{\ge 0}\to\mathbb{R}_{\ge 0}$ is a monotone scalar transform specified in Section~\ref{SE5}; the raw distance encoding corresponds to $\psi=\mathrm{id}$. The target Laplacian spectral encoding is the truncated coordinate $\Phi_t^{(m)}(v)\in\mathbb{R}^m$ (or a sign-/basis-invariant variant thereof).

Our central question is:

\textit{Under suitable structural conditions on $G$ and a well-posed anchor configuration $S$, can one construct an explicit algebraic map that takes $\tilde{\zeta}(v\mid S)$ as input and outputs $\Phi_t^{(m)}(v)$ with quantitatively controlled error, and get a bound associated matrix-level discrepancy (e.g., in Frobenius norm) between the induced distance/kernel quantities?}

In Section~\ref{SE5} we formalize these conditions and show that such a linkage exists, with error controlled (up to explicit constants) by a structural mismatch parameter $\delta$, in a logarithmic neighborhood regime consistent with common locality scales in prior analyses.

\section{Theoretical Analysis}
\label{SE5}

This section analyzes the algebraic relation between shortest-path distance encodings (DE) and truncated Laplacian spectral coordinates (LapPE) on $r$-regular graphs. All proofs are deferred to Appendix~\ref{app:proofs}. Throughout this section we use the (symmetric normalized) Laplacian in the $r$-regular form $L = I-\frac{1}{r}A,$
so that the diffusion operator is $K_t=e^{-tL}$ and the truncated diffusion-map embedding is $\Phi_t^{(m)}$ as defined in Section~\ref{SE3}.

\subsection{Assumptions}

\begin{assumption}
\label{ass:Gnr_main}
For the theoretical analysis, we consider graphs drawn from the random regular model $G \sim \mathcal{G}_{n,r}$, the uniform distribution on labeled $r$-regular graphs on $[n]$ with fixed degree $r \ge 3$.
\end{assumption}

\begin{assumption}
\label{ass:inj_main}
There exist $m \in \mathbb{N}$ and $\alpha > 0$ such that, with high probability over $G \sim \mathcal{G}_{n,r}$, the truncated embedding $\Phi_t^{(m)} : V \to \mathbb{R}^m$ is injective and satisfies
\begin{equation}
 \min_{u \neq v} \bigl\| \Phi_t^{(m)}(u) - \Phi_t^{(m)}(v) \bigr\|_2
 \;\ge\; n^{-\alpha}.
\end{equation}
\end{assumption}

These assumptions hold with high probability for $G\sim \mathcal{G}_{n,r}$ in the same logarithmic regime commonly used in analyses of Laplacian and distance-based positional encodings.

\subsection{Main Theoretical Results}

We first define an explicit trilateration operator that maps distance features to an $m$-dimensional spectral coordinate.

\begin{definition}
\label{def:trilateration_operator}
Fix $t>0$ and $m \in \mathbb{N}$, and let $a_1,\dots,a_{m+1} \in V$ be anchors with
\begin{equation}
 p_i = \Phi_t^{(m)}(a_i) \in \mathbb{R}^m, \quad i = 1,\dots,m+1.
\end{equation}
Define
\begin{equation}
 A
 = 2
 \begin{pmatrix}
   (p_1 - p_{m+1})^\top \\
   \vdots \\
   (p_m - p_{m+1})^\top
 \end{pmatrix}
 \in \mathbb{R}^{m\times m}.
\end{equation}
For any $r = (r_1,\dots,r_{m+1})^\top \in \mathbb{R}^{m+1}$, define
\begin{equation}
 b(r)
 =
 \begin{pmatrix}
   \|p_1\|_2^2 - \|p_{m+1}\|_2^2 + r_{m+1}^{\,2} - r_1^{\,2} \\
   \vdots \\
   \|p_m\|_2^2 - \|p_{m+1}\|_2^2 + r_{m+1}^{\,2} - r_m^{\,2}
 \end{pmatrix}
 \in \mathbb{R}^m.
\end{equation}
Let $\psi : [0,R] \to \mathbb{R}_+$ be the strictly increasing scalar function from Theorem~\ref{thm:local_monotone_link_pro} and write $\psi_*(y)$ for its elementwise application to $y\in\mathbb{R}^{m+1}$, i.e.,
\begin{equation}
 \bigl(\psi_*(y)\bigr)_i := \psi(y_i),\quad i=1,\dots,m+1.
\end{equation}
Recall the (raw) distance encoding to anchors $S=\{a_1,\dots,a_{m+1}\}$:
\begin{equation}
\resizebox{\columnwidth}{!}{$
 \zeta(v \mid S)
 = \bigl( \SPD(a_1,v), \dots, \SPD(a_{m+1},v) \bigr)^\top \in \mathbb{R}^{m+1}.
 $}
\end{equation}
Whenever $A$ is invertible, define
\begin{equation}
 T(v)
 := A^{-1} b\bigl( \psi_*(\zeta(v \mid S)) \bigr)
 \in \mathbb{R}^m.
\end{equation}
\end{definition}

\paragraph{Auxiliary quantities.}
Let $\{(\lambda_j,\varphi_j)\}_{j=1}^n$ be an orthonormal eigendecomposition of $L$. For $m\in\mathbb{N}$ and $t>0$, define the truncation tail
\begin{equation}\label{eq:tail_def}
\Tail_t^{(m)}(u,v)^2
:=\sum_{j=m+2}^{n} e^{-2t\lambda_j}\bigl(\varphi_j(u)-\varphi_j(v)\bigr)^2.
\end{equation}
Let $p:V\to\mathbb{R}^m$ and $V_1,\dots,V_{m+1}\stackrel{\mathrm{i.i.d.}}{\sim}\mathrm{Unif}(V)$. Define the degeneracy probability
\begin{equation}
\label{eq:eta_def}
\resizebox{\columnwidth}{!}{$
\eta_n \coloneqq
\mathbb{P}\!\left(
\det\!\bigl[p(V_1)-p(V_{m+1})\ \ \cdots\ \ p(V_m)-p(V_{m+1})\bigr]=0
\right).
$}
\end{equation}

\begin{theorem}
\label{thm:local_monotone_link_pro}
Fix $t>0$ and $m\in\mathbb{N}$. Let $G=(V,E)$ be a finite connected graph and let $R\ge 1$.
Let $\psi:[0,R]\to\mathbb{R}_+$ be strictly increasing. Define
\begin{equation}\label{eq:Dgeom_def}
\begin{aligned}
&\Dgeom(G;R,t,\psi)\\
:=&\sup_{\SPD(u,v)\le R}\bigl|d_t(u,v)-\psi(\SPD(u,v))\bigr|
\end{aligned}
\end{equation}
and
\begin{equation}\label{eq:Dtail_def}
\Dtail(G;R,t,m)
:=\sup_{\SPD(u,v)\le R}\Tail_t^{(m)}(u,v).
\end{equation}
Then, for all $u,v\in V$ with $\SPD(u,v)\le R$,
\begin{equation}\label{eq:main_bound}
\begin{aligned}
\bigl|d_t^{(m)}(u,v)-\psi(\SPD(u,v))\bigr|
&\le \Dgeom(G;R,t,\psi)\\&+\Dtail(G;R,t,m).
\end{aligned}
\end{equation}
\end{theorem}

For later use, define the local linkage/truncation error level
\begin{equation}
\label{eq:deltaL_def}
\delta_n^L \;\coloneqq\; \Dgeom(G;R,t,\psi)\;+\;\Dtail(G;R,t,m).
\end{equation}

\begin{proposition}
\label{prop:generic_anchors}
Fix $m\ge 1$ and let $p:V\to\mathbb{R}^m$ be the truncated diffusion embedding
$p(v)\coloneqq \Phi_t^{(m)}(v)$. Draw anchors
$a_1,\dots,a_{m+1}\stackrel{\mathrm{i.i.d.}}{\sim}\mathrm{Unif}(V)$ and set $p_i\coloneqq p(a_i)$.
Define
\begin{equation}\label{eq:anchor_matrix}
M \coloneqq \bigl[p_1-p_{m+1}\ \ \cdots\ \ p_m-p_{m+1}\bigr]\in\mathbb{R}^{m\times m}.
\end{equation}
Then $p_1,\dots,p_{m+1}$ are affinely independent in $\mathbb{R}^m$ if and only if $\det(M)\neq 0$.
Moreover, the trilateration matrix in Definition~\ref{def:trilateration_operator} satisfies $A=2M^\top$.

Assume the degeneracy probability $\eta_n$ in \eqref{eq:eta_def} satisfies $\eta_n=o(1)$.
Then $\mathbb{P}\bigl(\det(M)\neq 0\bigr)=1-\eta_n=1-o(1)$.
\end{proposition}

\begin{remark}[Jittered anchors are generic a.s.]
Fix $m\ge 1$ and let $p_1,\dots,p_{m+1}\in\mathbb{R}^m$ be arbitrary points.
Let $\varepsilon>0$ and let $\xi_1,\dots,\xi_{m+1}\in\mathbb{R}^m$ be random vectors whose joint law is absolutely
continuous with respect to Lebesgue measure. Setting $\tilde p_i\coloneqq p_i+\varepsilon\xi_i$ and
\[
\tilde M \coloneqq \bigl[\tilde p_1-\tilde p_{m+1}\ \ \cdots\ \ \tilde p_m-\tilde p_{m+1}\bigr],
\]
we have $\mathbb{P}\bigl(\det(\tilde M)\neq 0\bigr)=1$.
\end{remark}

The next two theorems bound the pointwise reconstruction error $\|\Phi_t^{(m)}(v)-T(v)\|_2$ and the induced matrix-level discrepancy.

\begin{theorem}
\label{thm:DE_to_LapPE_pointwise}
Let $G \sim \mathcal{G}_{n,r}$ with fixed $r \ge 3$, and fix $t>0$ and $m \in \mathbb{N}$.
Let $a_1,\dots,a_{m+1}$ be anchors satisfying Proposition~\ref{prop:generic_anchors} (and, if needed, the jittered-genericity remark above), and
let $\Phi_t^{(m)}$ be the truncated spectral embedding. Suppose
Assumptions~\ref{ass:Gnr_main}-\ref{ass:inj_main} hold, and Theorem~\ref{thm:local_monotone_link_pro}
holds for $d_t^{(m)}$ with error $\delta_n^L$.

For a node $v\in V$ such that $\SPD(a_i,v)\le R$ for all $i$, define
\begin{equation}
r_i := \psi\bigl(\SPD(a_i,v)\bigr),\quad
r_i^\star := d_t^{(m)}(a_i,v).
\end{equation}
Then, with probability $1-o(1)$ over $G$, we have
\begin{equation}
\resizebox{\columnwidth}{!}{$
\bigl\|\Phi_t^{(m)}(v)-T(v)\bigr\|_2
\;\le\;
\|A^{-1}\|_{\mathrm{op}}\;\sqrt{m}\,\Bigl(4\,\rho_R\,\delta_n^L + 2(\delta_n^L)^2\Bigr),
$}
\end{equation}
where
\begin{equation}
\rho_R := \max_{0\le d\le R}\psi(d)\;+\;\delta_n^L.
\end{equation}
In particular, for $\delta_n^L\le 1$,
\begin{equation}
\bigl\|\Phi_t^{(m)}(v)-T(v)\bigr\|_2
\;\le\;
\|A^{-1}\|_{\mathrm{op}}\;\sqrt{m}\,\bigl(4\rho_R+2\bigr)\,\delta_n^L.
\end{equation}
\end{theorem}

\begin{theorem}
\label{thm:DE_to_LapPE_matrix}
Under the same setting and assumptions as in Theorem~\ref{thm:DE_to_LapPE_pointwise},
define
\begin{equation}
(D_{\mathrm{SPD}})_{v,i}=\SPD(v,a_i),
(D_{\mathrm{diff}}^{(m)})_{v,i}=d_t^{(m)}(v,a_i),
\end{equation}
and let $\psi_*(D_{\mathrm{SPD}})$ denote the elementwise application of $\psi$.
Then, with probability $1-o(1)$,
\begin{equation}
\bigl\|D_{\mathrm{diff}}^{(m)}-\psi_*(D_{\mathrm{SPD}})\bigr\|_F
\;\le\;
\delta_n^L\,\sqrt{n(m+1)}.
\end{equation}
In particular, the average per-entry discrepancy is at most $\delta_n^L$.
\end{theorem}

Proofs are given in Appendix~\ref{app:proofs}. Theorem~\ref{thm:DE_to_LapPE_pointwise} combines the local monotone linkage (Theorem~\ref{thm:local_monotone_link_pro}) with a perturbation analysis of the linear system in Definition~\ref{def:trilateration_operator}, and Theorem~\ref{thm:DE_to_LapPE_matrix} follows by applying the linkage uniformly to all node-anchor pairs and summing the resulting errors.

\subsection{Implications for expressivity relative to NoPE}

We recall a standard consequence of distance encodings for message passing GNNs on random regular graphs.

\begin{theorem}
\label{thm:de-wl}
Let $G \sim \mathcal{G}_{n,r}$ be drawn from the random $r$-regular graph model with fixed $r \ge 3$, and let $\mathcal{F}_{\mathrm{MP}}$ denote the class of $T$-layer message passing GNNs without positional encodings (NoPE), whose distinguishing power is upper bounded by the 1-WL test \citeyearpar{morris2019weisfeiler,xu2018powerful}. Let $\mathcal{F}_{\mathrm{DE}}$ denote the class of $T$-layer message passing GNNs augmented with a distance encoding based on shortest-path distances to $k = \Theta(\log n)$ anchors, as in \citeyearpar{li2009distance}. Then there exists a family of node-level classification tasks on $\mathcal{G}_{n,r}$ such that, with high probability over $G$,
\begin{enumerate}
   \item no NoPE message passing GNN in $\mathcal{F}_{\mathrm{MP}}$ can realize the target labels (because 1-WL cannot distinguish the relevant nodes); but
   \item some DE-augmented GNN in $\mathcal{F}_{\mathrm{DE}}$ separates all label classes exactly.
\end{enumerate}
In particular, on random regular graphs the function class $\mathcal{F}_{\mathrm{DE}}$ is strictly more expressive than the NoPE class $\mathcal{F}_{\mathrm{MP}}$.
\end{theorem}

\begin{corollary}
\label{cor:de-vs-nope}
Under Assumptions~\ref{ass:Gnr_main} and \ref{ass:inj_main} (and the same random-regular/diffusion-geometry regime used above), consider the NoPE and DE variants of our backbone architecture: (i) $\mathcal{F}_{\mathrm{NoPE}}$, the backbone without positional encodings; and (ii) $\mathcal{F}_{\mathrm{DE}}$, the same backbone augmented with $\zeta(\cdot\mid S)$ using $k=\Theta(\log n)$ anchors. Then, with high probability over $G\sim\mathcal{G}_{n,r}$ and random anchors $S$, $\mathcal{F}_{\mathrm{DE}}$ is strictly more expressive than $\mathcal{F}_{\mathrm{NoPE}}$; in particular, there exist node classification tasks realizable by some DE-augmented instance but by no NoPE instance.
\end{corollary}

Corollary~\ref{cor:de-vs-nope} follows from Theorem~\ref{thm:de-wl} and the equivalence between NoPE message passing GNNs and the 1-WL test \citeyearpar{morris2019weisfeiler,xu2018powerful}. Together with Theorems~\ref{thm:DE_to_LapPE_pointwise}-\ref{thm:DE_to_LapPE_matrix}, this yields the expressivity chain on random regular graphs
\begin{equation}
   \text{NoPE} \;\subsetneq\; \text{DE-augmented} \;\approx\; \text{LapPE-augmented},
\end{equation}
where $\approx$ is in the sense of the reconstruction and Frobenius-gap bounds above.

\section{Experimental Setup}
\label{SE6}
We conduct a controlled comparison of positional encodings under a unified GNP-based DDI prediction backbone. Unless stated otherwise, all implementation and hyperparameter details follow \citeyearpar{yan2025multiscalegraphneuralprocess} and are provided in Appendix~B.

\noindent\textbf{Datasets and protocol.}
We evaluate on DrugBank and ChCh-Miner, where nodes are drugs and edges are known interactions. Molecular graphs are built from SMILES using RDKit. We adopt the inductive link prediction protocol in baselines with train/val/test splits over drug pairs.

\noindent\textbf{Baselines.}
To isolate positional effects, we fix the backbone and compare three primary variants: NoPE, distance encoding (DE), and Laplacian positional encoding (LapPE). We additionally report RWSE and heat-kernel signatures (HKS) as reference positional/structural baselines under the same backbone and training protocol.

\noindent\textbf{Model configuration.}
All models are implemented in PyTorch and PyTorch Geometric with standard atom/bond features. LapPE uses the first $m$ non-trivial eigenvectors of the normalized Laplacian; DE uses shortest-path distances to $k$ anchors with a radial transform $\psi(\cdot)$. RWSE uses return probabilities at steps $\{1,2,4,8,16\}$ and HKS uses diffusion times $\{0.1,0.5,1,2,5\}$ from a truncated Laplacian spectrum. Architecture, normalization, and optimization settings are in Appendix~B.

\noindent\textbf{Evaluation.}
DDI prediction is treated as binary classification; we report AUROC and F1 on the held-out test set, with the F1 threshold selected on the validation set. Primary results (NoPE/DE/LapPE) are averaged over three random seeds; RWSE/HKS are reported using the available runs.

\noindent\textbf{Ablations.}
We probe the DE design by varying $\psi(\cdot)$ on DrugBank with the number of anchors fixed, and varying the number of anchors $k$ on ChCh-Miner with $\psi(\cdot)$ fixed. All ablations reuse the main preprocessing and training settings; detailed results are in Section~\ref{subsec:de-ablation-results}.

\section{Experimental Results and Analysis}
\label{SE7}
We evaluate the proposed DE-LapPE bridge from three perspectives: (i) theory-aligned validation on random $r$-regular graphs (Section~\ref{SE5}); (ii) controlled spectral-approximation on real DrugBank molecular graphs; and (iii) downstream DDI prediction under a fixed GNP backbone (NoPE/DE/LapPE), together with targeted DE ablations on $\psi(\cdot)$ and the number of anchors $k$.

\subsection{Theory-aligned validation on random $r$-regular graphs}
We run a controlled study on $G\sim \mathcal{G}_{n,r}$ with $r=6$ and $n\in\{256,512,1024,2048\}$ (three seeds per $n$). For each graph, we form $L=I-\frac{1}{r}A$ and compute truncated diffusion coordinates $\Phi_t^{(m)}$ with $t=1.0$ and $m=8$. We set the locality radius $R=\lceil \log n\rceil$ and fit a monotone map $\psi$ from $\SPD$ to $d_t^{(m)}$ using isotonic regression on pairs with $\SPD(u,v)\le R$.

We report (i) the empirical linkage error $\hat\delta_L^n \coloneqq \max_{\SPD(u,v)\le R}|d_t^{(m)}(u,v)-\psi(\SPD(u,v))|$; (ii) the normalized Frobenius gap $\|D_{\mathrm{diff}}^{(m)}-\psi_*(D_{\mathrm{SPD}})\|_F/\sqrt{n(m+1)}$ on node-anchor matrices; and (iii) trilateration reconstruction with $m{+}1$ anchors via $T(v)$, summarized by the median pointwise error $\|\Phi_t^{(m)}(v)-T(v)\|_2$ and the median $\mathrm{cond}(A)$.

Table~\ref{tab:rrg_validation} shows that both $\hat\delta_L^n$ and the normalized Frobenius gap decrease as $n$ grows, and trilateration reconstruction errors remain small in typical cases.

\begin{table}[t]
\centering
\caption{Validation on random $r$-regular graphs ($r{=}6$, $t{=}1.0$, $m{=}8$; three seeds).}
\label{tab:rrg_validation}
\small
\textbf{(a) Linkage and matrix-level discrepancy.}\\[1mm]
\begin{tabular}{c c c c}
\hline
$n$ & $R$ & $\hat\delta_L^n$ (mean$\pm$std) & Frob-gap (mean$\pm$std) \\
\hline
256  & 6 & $0.1580\pm0.0035$ & $0.0350\pm0.0009$ \\
512  & 7 & $0.1099\pm0.0051$ & $0.0290\pm0.0028$ \\
1024 & 7 & $0.0855\pm0.0040$ & $0.0217\pm0.0014$ \\
2048 & 8 & $0.0664\pm0.0024$ & $0.0156\pm0.0002$ \\
\hline
\end{tabular}
\vspace{3mm}
\textbf{(b) Pointwise reconstruction and conditioning.}\\[1mm]
\begin{tabular}{c c c c}
\hline
$n$ & $R$ & $\|\Phi-T\|_2$ (median) & cond$(A)$ (median) \\
\hline
256  & 6 & $0.449$ & $46.89$ \\
512  & 7 & $0.757$ & $112.41$ \\
1024 & 7 & $0.181$ & $63.05$ \\
2048 & 8 & $0.0845$ & $30.00$ \\
\hline
\end{tabular}
\end{table}

\subsection{Spectral approximation of diffusion geometry on DrugBank}
\label{subsec:diffusion-approx}

We analyze 80 DrugBank molecular graphs (15-200 nodes). For each graph, we compute the diffusion kernel ($t=1$) and diffusion map (top $m=8$ components), and approximate the kernel via an anchor-based DE Nystr\"om scheme with $k=32$ anchors (farthest-point sampling) and Tikhonov regularization.

\begin{figure}[htbp]
  \centering
  \includegraphics[width=\columnwidth]{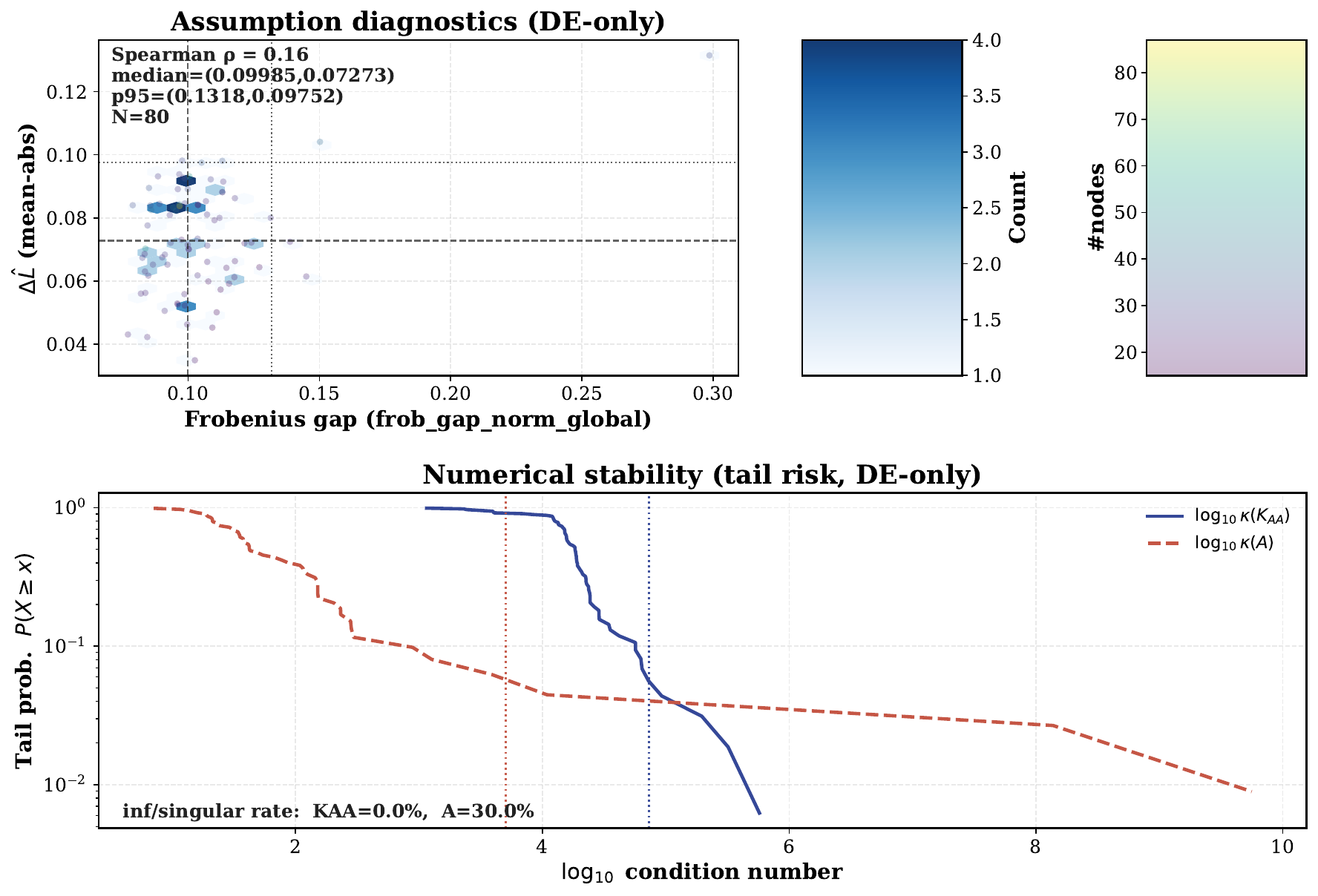}
  \caption{\textbf{DrugBank diffusion-geometry diagnostics (80 graphs).}
  \textbf{Top:} normalized Frobenius gap vs.\ monotonicity residual $\Delta\hat{L}$ (median/p95 and Spearman $\rho$).
  \textbf{Bottom:} tail CCDF of $\log_{10}\kappa(K_{AA})$ and $\log_{10}\kappa(A)$.}
  \label{fig:drugbank_diagnostics}
  \vspace{-5mm}
\end{figure}

Nystr\"om-DE is accurate: relative kernel Frobenius error $0.024\pm0.021$ (median $0.020$), coordinate MSE $3.9\times10^{-4}$ after Procrustes alignment, and mean absolute Pearson correlation $0.988$ between approximate and exact diffusion distances. Figure~\ref{fig:drugbank_diagnostics} further reports the fitted local isotonic link $\psi^*$ (with $R=\lceil\log n\rceil$) and conditioning diagnostics, showing small linkage/gap values and heavy-tailed condition numbers that motivate regularization.

\subsection{Effect of positional encodings on DDI prediction}

Under the same GNP-DDI backbone, training schedule, and data splits, we compare NoPE, DE, LapPE, and two reference baselines (RWSE, HKS); only the positional encoding module is changed. Table~\ref{tab:pe_all} reports test AUROC and F1 (mean$\pm$std over three runs).

LapPE is best overall. On DrugBank, positional information is critical: NoPE attains $0.890/0.820$ (AUROC/F1), DE improves to $0.976/0.927$, and LapPE further to $0.980/0.934$. On ChCh-Miner, performance is already strong, but LapPE still yields consistent gains ($0.946/0.879$) over NoPE ($0.938/0.870$) and DE ($0.938/0.869$). RWSE is competitive on ChCh-Miner, while HKS is weaker on DrugBank under this plug-in setting.\footnote{RWSE uses return probabilities at steps $\{1,2,4,8,16\}$; HKS uses diffusion times $\{0.1,0.5,1,2,5\}$ with truncated spectrum dimension $k{=}32$ (normalized Laplacian).}

\begin{table}[t]
  \centering
  \small
  \setlength{\tabcolsep}{3pt}
  \renewcommand{\arraystretch}{0.95}
  \caption{Test AUROC and F1 (mean $\pm$ std over three runs) on DrugBank and ChCh-Miner under the same GNP-DDI backbone and training protocol. Only the positional encoding is varied.}
  \label{tab:pe_all}
  \begin{tabular}{@{}l l cc@{}}
    \toprule
    Dataset & Method & Test AUROC & Test F1 \\
    \midrule
    \multirow{5}{*}{DrugBank}
      & NoPE  & 0.890\,$\pm$\,0.002 & 0.820\,$\pm$\,0.003 \\
      & DE    & 0.976\,$\pm$\,0.002 & 0.927\,$\pm$\,0.004 \\
      & LapPE & 0.980\,$\pm$\,0.003 & 0.934\,$\pm$\,0.006 \\
      \cmidrule(lr){2-4}
      & RWSE  & 0.892\,$\pm$\,0.008 & 0.812\,$\pm$\,0.003 \\
      & HKS   & 0.863\,$\pm$\,0.013 & 0.787\,$\pm$\,0.007 \\
    \midrule
    \multirow{5}{*}{ChCh-Miner}
      & NoPE  & 0.938\,$\pm$\,0.003 & 0.870\,$\pm$\,0.002 \\
      & DE    & 0.938\,$\pm$\,0.006 & 0.869\,$\pm$\,0.004 \\
      & LapPE & 0.946\,$\pm$\,0.002 & 0.879\,$\pm$\,0.003 \\
      \cmidrule(lr){2-4}
      & RWSE  & 0.944\,$\pm$\,0.003 & 0.876\,$\pm$\,0.002 \\
      & HKS   & 0.941\,$\pm$\,0.002 & 0.870\,$\pm$\,0.004 \\
    \bottomrule
  \end{tabular}
\end{table}

\subsection{Distance Encoding Ablation Results}
\label{subsec:de-ablation-results}

\begin{table}[t]
  \centering
  \small
  \setlength{\tabcolsep}{2.5pt}
  \renewcommand{\arraystretch}{0.95}
  \caption{Ablation study of distance encoding (DE) on DrugBank and ChCh-Miner. The best configuration for each dataset is shaded.}
  \label{tab:de_ablation}

  \textbf{(a) DrugBank ($k=16$, varying $\psi(\cdot)$)}\\[1pt]
  \begin{tabular}{@{}lccccc@{}}
    \toprule
    $\psi$ & Par. & Val AUC & Val F1 & Test AUC & Test F1 \\
    \midrule
    identity & 1.168 & 0.9430 & 0.8795 & 0.9451 & 0.8835 \\
    \rowcolor{gray!15}
    exp      & 1.168 & 0.9762 & 0.9279 & 0.9760 & 0.9281 \\
    log1p    & 1.168 & 0.9723 & 0.9196 & 0.9721 & 0.9207 \\
    \bottomrule
  \end{tabular}

  \vspace{4pt}

  \textbf{(b) ChCh-Miner ($\psi(d)=\exp(-d)$, varying $k$)}\\[1pt]
  \begin{tabular}{@{}rccccc@{}}
    \toprule
    $k$ & Par. & Val AUC & Val F1 & Test AUC & Test F1 \\
    \midrule
     4 & 0.123 & 0.9415 & 0.8715 & 0.9410 & 0.8718 \\
     8 & 0.123 & 0.9411 & 0.8715 & 0.9411 & 0.8727 \\
    16 & 0.123 & 0.9406 & 0.8709 & 0.9412 & 0.8722 \\
    \rowcolor{gray!15}
    32 & 0.124 & 0.9455 & 0.8778 & 0.9453 & 0.8784 \\
    \bottomrule
  \end{tabular}
\end{table}

On DrugBank (Table~\ref{tab:de_ablation}a), with $k=16$ anchors, the choice of $\psi(\cdot)$ is decisive: identity mapping underperforms (test $0.9451/0.8835$ AUROC/F1), while $\psi=\exp$ performs best (test $0.9760/0.9281$), and $\log(1{+}d)$ is slightly worse. Parameter counts are identical (1.168M), so the gains come from the distance shaping rather than capacity.

On ChCh-Miner (Table~\ref{tab:de_ablation}b), fixing $\psi(d)=\exp(-d)$, performance is stable for $k\in\{4,8,16\}$ (test $\approx 0.941/0.872$) and improves at $k=32$ (test $0.9453/0.8784$) with negligible parameter increase (0.123M$\to$0.124M).

\subsection{Qualitative case study on a DrugBank molecular graph}

\begin{figure}[!htbp]
  \centering
  \begin{subfigure}[t]{0.48\columnwidth}
    \centering
    \includegraphics[width=\linewidth]{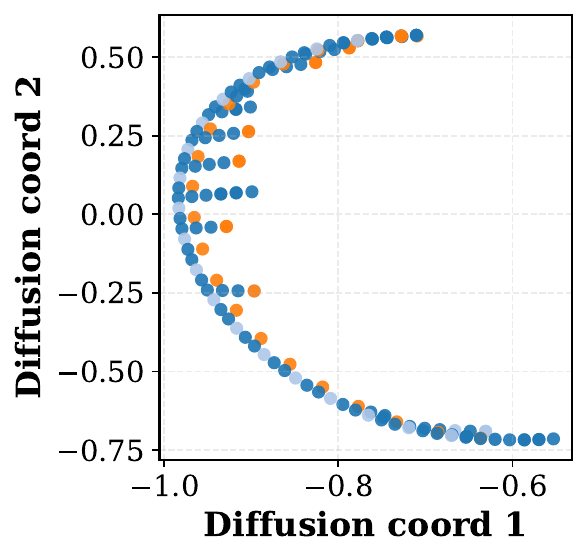}
    \caption{Full Gaussian kernel eigen-decomposition}
  \end{subfigure}
  \hfill
  \begin{subfigure}[t]{0.48\columnwidth}
    \centering
    \includegraphics[width=\linewidth]{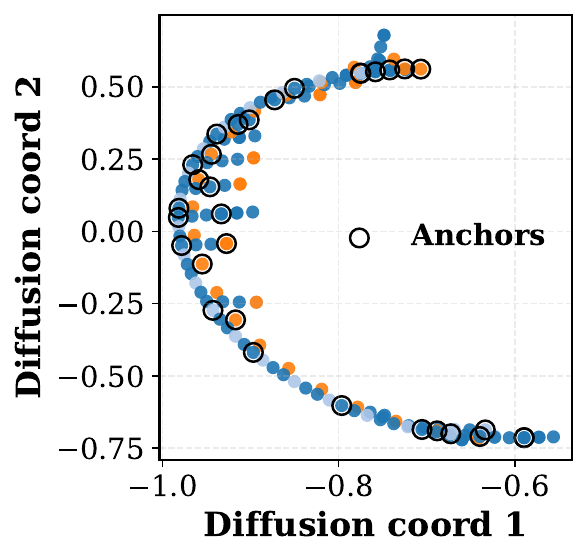}
    \caption{DE-based Nystr\"om embedding}
  \end{subfigure}
  \vspace{0.6em}
  \begin{subfigure}[t]{0.75\columnwidth}
    \centering
    \includegraphics[width=\linewidth]{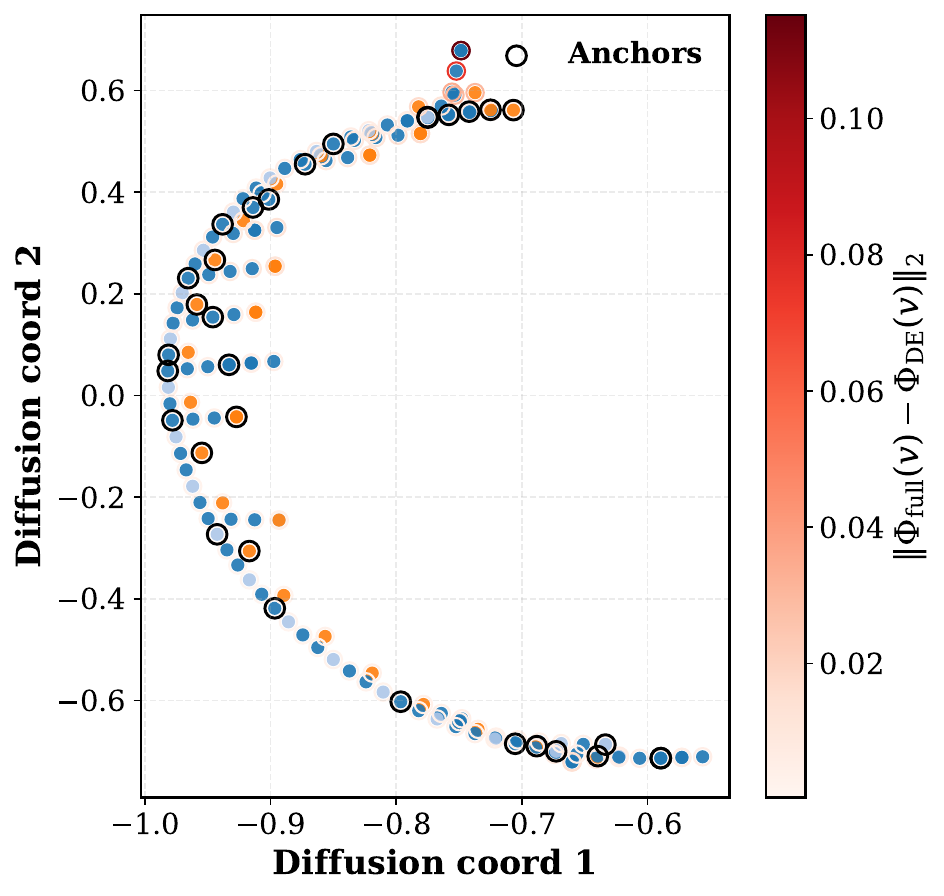}
    \caption{Node-wise embedding error on DE Nystr\"om}
  \end{subfigure}
  \caption{Qualitative comparison of diffusion-based embeddings for a single DrugBank molecular graph (DB00006).}
  \label{fig:case_study_drugbank}
\end{figure}

We visualize a single DrugBank molecule (DB00006, 155 atoms) to compare the reference diffusion-map embedding from the full Gaussian-kernel eigendecomposition with the DE-based Nystr\"om embedding using $k=32$ anchors (after Procrustes alignment). As shown in Fig.~\ref{fig:case_study_drugbank}(a-b), DE Nystr\"om closely matches the global diffusion geometry.

Fig.~\ref{fig:case_study_drugbank}(c) reports the node-wise error $\|\Phi_{\mathrm{full}}(v)-\Phi_{\mathrm{DE}}(v)\|_2$: most nodes are at $10^{-2}$ scale or below, with mean $7.8\times10^{-3}$ and maximum $1.15\times10^{-1}$. Together with the aggregate results in Subsection~\ref{subsec:diffusion-approx}, this supports that a modest anchor set can recover leading diffusion coordinates with high fidelity in practice.
\FloatBarrier

\section*{Limitations}

We note a few limitations and practical considerations of our theory and experiments.

\textbf{Theory scope.}
Our guarantees are established under a random regular graph model and a local linkage between diffusion and shortest-path distances. They do not directly cover graphs with strong degree heterogeneity, pronounced community structure, edge weights, or directionality that commonly arise in practice.

\textbf{Approximation design choices.}
The dependence of the error on truncation level, anchor placement, and the radial transform is only partially characterized. We do not analyze data-driven anchor selection, learned distance metrics, or learned (potentially non-monotone) distance transforms beyond the targeted ablations.

\textbf{Empirical generality.}
We evaluate on two DDI datasets and a single multi-scale Graph Neural Process backbone; results may not fully transfer to other domains, datasets with different labeling/curation mechanisms, or architectures.

\textbf{Data and evaluation caveats.}
DDI resources can be incomplete and subject to reporting/curation biases, and unobserved drug pairs used as negatives may include false negatives. As a result, performance under standard splits may not fully reflect real-world pharmacovigilance settings.

\textbf{Potential risks.}
Our DDI prediction experiments are intended for research and benchmarking rather than clinical decision making. Misuse or over-reliance on predictions (false positives/negatives) could lead to inappropriate conclusions without expert review and external validation.

\textbf{Future directions.}
Positional encodings are treated as fixed precomputed features; we leave joint end-to-end learning of anchors, radial maps, and spectral regularizers to future work.

\textbf{Use of AI assistants.}
We used AI assistants to support code development (e.g., debugging and boilerplate). All experimental results, analyses, and claims were produced and verified by the authors.

\section*{Acknowledgments}
The authors would like to express their sincere gratitude to all the anonymous reviewers for their careful reading and insightful suggestions.


\bibliography{custom}

@article{bronstein2017geometric,
  title={Geometric deep learning: going beyond euclidean data},
  author={Bronstein, Michael M and Bruna, Joan and LeCun, Yann and Szlam, Arthur and Vandergheynst, Pierre},
  journal={IEEE Signal Processing Magazine},
  volume={34},
  number={4},
  pages={18--42},
  year={2017},
  publisher={IEEE}
}

@article{zhou2020graph,
  title={Graph neural networks: A review of methods and applications},
  author={Zhou, Jie and Cui, Ganqu and Hu, Shengding and Zhang, Zhengyan and Yang, Cheng and Liu, Zhiyuan and Wang, Lifeng and Li, Changcheng and Sun, Maosong},
  journal={AI open},
  volume={1},
  pages={57--81},
  year={2020},
  publisher={Elsevier}
}

@article{wu2020comprehensive,
  title={A comprehensive survey on graph neural networks},
  author={Wu, Zonghan and Pan, Shirui and Chen, Fengwen and Long, Guodong and Zhang, Chengqi and Yu, Philip S},
  journal={IEEE transactions on neural networks and learning systems},
  volume={32},
  number={1},
  pages={4--24},
  year={2020},
  publisher={IEEE}
}

@inproceedings{gilmer2017neural,
  title={Neural message passing for quantum chemistry},
  author={Gilmer, Justin and Schoenholz, Samuel S and Riley, Patrick F and Vinyals, Oriol and Dahl, George E},
  booktitle={International conference on machine learning},
  pages={1263--1272},
  year={2017},
  organization={Pmlr}
}

@article{yan2025resolvingnodeidentifiabilitygraph,
  title={Resolving Node Identifiability in Graph Neural Processes via Laplacian Spectral Encodings},
  author={Yan, Zimo and Xie, Zheng and Liu, Chang and Wang, Yuan},
  journal={arXiv preprint arXiv:2511.19037},
  year={2025}
}

@article{belkin2003laplacian,
  title={Laplacian eigenmaps for dimensionality reduction and data representation},
  author={Belkin, Mikhail and Niyogi, Partha},
  journal={Neural computation},
  volume={15},
  number={6},
  pages={1373--1396},
  year={2003},
  publisher={MIT Press}
}

@article{vonluxburg2007tutorial,
  title={A tutorial on spectral clustering},
  author={Von Luxburg, Ulrike},
  journal={Statistics and computing},
  volume={17},
  number={4},
  pages={395--416},
  year={2007},
  publisher={Springer}
}

@article{coifman2006diffusion,
  title={Diffusion maps},
  author={Coifman, Ronald R and Lafon, St{\'e}phane},
  journal={Applied and computational harmonic analysis},
  volume={21},
  number={1},
  pages={5--30},
  year={2006},
  publisher={Elsevier}
}

@article{coifman2005geometric,
  title={Geometric diffusions as a tool for harmonic analysis and structure definition of data: Diffusion maps},
  author={Coifman, Ronald R and Lafon, Stephane and Lee, Ann B and Maggioni, Mauro and Nadler, Boaz and Warner, Frederick and Zucker, Steven W},
  journal={Proceedings of the national academy of sciences},
  volume={102},
  number={21},
  pages={7426--7431},
  year={2005},
  publisher={National Academy of Sciences}
}

@article{dwivedi2023benchmarking,
  title={Benchmarking graph neural networks},
  author={Dwivedi, Vijay Prakash and Joshi, Chaitanya K and Luu, Anh Tuan and Laurent, Thomas and Bengio, Yoshua and Bresson, Xavier},
  journal={Journal of Machine Learning Research},
  volume={24},
  number={43},
  pages={1--48},
  year={2023}
}

@article{maskey2022generalized,
  title={Generalized laplacian positional encoding for graph representation learning},
  author={Maskey, Sohir and Parviz, Ali and Thiessen, Maximilian and St{\"a}rk, Hannes and Sadikaj, Ylli and Maron, Haggai},
  journal={arXiv preprint arXiv:2210.15956},
  year={2022}
}

@inproceedings{eliasof2023graph,
  title={Graph positional encoding via random feature propagation},
  author={Eliasof, Moshe and Frasca, Fabrizio and Bevilacqua, Beatrice and Treister, Eran and Chechik, Gal and Maron, Haggai},
  booktitle={International conference on machine learning},
  pages={9202--9223},
  year={2023},
  organization={PMLR}
}

@article{li2009distance,
  title={Distance encoding-design provably more powerful graph neural networks for structural representation learning},
  author={Li, Pan and Wang, Yanbang and Wang, Hongwei and Leskovec, Jure},
  journal={CoRR},
  pages={00142--00142},
  year={2009}
}

@article{zitnik2018modeling,
  title={Modeling polypharmacy side effects with graph convolutional networks},
  author={Zitnik, Marinka and Agrawal, Monica and Leskovec, Jure},
  journal={Bioinformatics},
  volume={34},
  number={13},
  pages={i457--i466},
  year={2018},
  publisher={Oxford University Press}
}

@article{ryu2018deep,
  title={Deep learning improves prediction of drug--drug and drug--food interactions},
  author={Ryu, Jae Yong and Kim, Hyun Uk and Lee, Sang Yup},
  journal={Proceedings of the national academy of sciences},
  volume={115},
  number={18},
  pages={E4304--E4311},
  year={2018},
  publisher={National Academy of Sciences}
}

@article{yan2025metamolgenneuralgraphmotif,
  title={MetaMolGen: A Neural Graph Motif Generation Model for De Novo Molecular Design},
  author={Yan, Zimo and Zhang, Jie and Xie, Zheng and Liu, Chang and Liu, Yizhen and Song, Yiping},
  journal={arXiv preprint arXiv:2504.15587},
  year={2025}
}

@inproceedings{lin2020kgnn,
  title={KGNN: Knowledge graph neural network for drug-drug interaction prediction.},
  author={Lin, Xuan and Quan, Zhe and Wang, Zhi-Jie and Ma, Tengfei and Zeng, Xiangxiang},
  booktitle={IJCAI},
  volume={380},
  pages={2739--2745},
  year={2020}
}

@article{deng2020multimodal,
  title={A multimodal deep learning framework for predicting drug--drug interaction events},
  author={Deng, Yifan and Xu, Xinran and Qiu, Yang and Xia, Jingbo and Zhang, Wen and Liu, Shichao},
  journal={Bioinformatics},
  volume={36},
  number={15},
  pages={4316--4322},
  year={2020},
  publisher={Oxford University Press}
}

@article{ma2023dual,
  title={A dual graph neural network for drug--drug interactions prediction based on molecular structure and interactions},
  author={Ma, Mei and Lei, Xiujuan},
  journal={PLOS Computational Biology},
  volume={19},
  number={1},
  pages={e1010812},
  year={2023},
  publisher={Public Library of Science San Francisco, CA USA}
}

@article{kipf2016semi,
  title={Semi-supervised classification with graph convolutional networks},
  author={Kipf, TN},
  journal={arXiv preprint arXiv:1609.02907},
  year={2016}
}

@inproceedings{morris2019weisfeiler,
  title={Weisfeiler and leman go neural: Higher-order graph neural networks},
  author={Morris, Christopher and Ritzert, Martin and Fey, Matthias and Hamilton, William L and Lenssen, Jan Eric and Rattan, Gaurav and Grohe, Martin},
  booktitle={Proceedings of the AAAI conference on artificial intelligence},
  volume={33},
  number={01},
  pages={4602--4609},
  year={2019}
}

@article{xu2018powerful,
  title={How powerful are graph neural networks?},
  author={Xu, Keyulu and Hu, Weihua and Leskovec, Jure and Jegelka, Stefanie},
  journal={arXiv preprint arXiv:1810.00826},
  year={2018}
}

@inproceedings{alon2021on,
  title        = {On the bottleneck of graph neural networks and its practical implications},
  author       = {Alon, Uri and Yahav, Eran},
  booktitle    = {International Conference on Learning Representations},
  year         = {2021}
}

@inproceedings{topping2021understanding,
  title        = {Understanding over-squashing and bottlenecks on graphs via curvature},
  author       = {Topping, James and Di Giovanni, Francesco and Chamberlain, Benjamin P and Dong, Xiaowen and Bronstein, Michael M},
  booktitle    = {International Conference on Learning Representations},
  year         = {2022}
}

@inproceedings{you2019pgnn,
  title        = {Position-aware graph neural networks for graph-level tasks},
  author       = {You, Jiaxuan and Ying, Rex and Ren, Xiang and Hamilton, William L and Leskovec, Jure},
  booktitle    = {Proceedings of the 36th International Conference on Machine Learning},
  pages        = {7134--7143},
  year         = {2019},
  organization = {PMLR},
  note         = {Graph-level variant of position-aware GNNs}
}

@inproceedings{ying2021transformers,
  title        = {Do transformers really perform badly for graph representation?},
  author       = {Ying, Zhitao and Cai, Tianjun and Luo, Shengjie and Zheng, Shuxin and Ke, Guolin and He, Di and Shen, Yanming and Liu, Tie-Yan},
  booktitle    = {Advances in Neural Information Processing Systems},
  volume       = {34},
  pages        = {28877--28888},
  year         = {2021}
}

@inproceedings{feng2020interpreter,
  title={Towards Interpretable Drug-Drug Interaction Prediction: A Graph-Based Approach with Molecular and Network-Level Explanations},
  author={Chen, Mengjie and Zhang, Ming and Qu, Cunquan},
  booktitle={Proceedings of the 31st ACM SIGKDD Conference on Knowledge Discovery and Data Mining V. 2},
  pages={203--214},
  year={2025}
}

@article{niu2024dasddi,
  title   = {DAS-DDI: A dual-view framework with drug association and similarity for drug--drug interaction prediction},
  author  = {Niu, Ding and Wang, Shuai and Li, Chunyu and others},
  journal = {Journal of Biomedical Informatics},
  year    = {2024},
  note    = {In press},
  doi     = {10.1016/j.jbi.2024.104547}
}

@article{wang2024structnetddi,
  title   = {StructNet-DDI: Molecular structure characterization-based representation learning for drug--drug interaction prediction},
  author  = {Wang, Jiarui and others},
  journal = {Molecules},
  volume  = {29},
  number  = {20},
  pages   = {4829},
  year    = {2024},
  publisher = {MDPI}
}

@article{he2022mffgnn,
  title   = {Multi-type feature fusion based on graph neural network for drug--drug interaction prediction},
  author  = {He, Chao and others},
  journal = {BMC Bioinformatics},
  volume  = {23},
  number  = {224},
  year    = {2022},
  publisher = {Springer}
}

@article{loshchilov2019decoupled,
  title   = {Decoupled weight decay regularization},
  author  = {Loshchilov, Ilya and Hutter, Frank},
  journal = {International Conference on Learning Representations},
  year    = {2019},
  note    = {ICLR}
}

@article{paszke2019pytorch,
  title   = {PyTorch: An Imperative Style, High-Performance Deep Learning Library},
  author  = {Paszke, Adam and Gross, Sam and Massa, Francisco and Lerer, Adam and Bradbury, James and Chanan, Gregory and Killeen, Trevor and Lin, Zeming and Gimelshein, Natalia and Antiga, Luca and others},
  journal = {Advances in Neural Information Processing Systems},
  volume  = {32},
  year    = {2019}
}

@article{fey2019fast,
  title   = {Fast Graph Representation Learning with PyTorch Geometric},
  author  = {Fey, Matthias and Lenssen, Jan Eric},
  journal = {Proceedings of the ICLR 2019 Workshop on Representation Learning on Graphs and Manifolds},
  year    = {2019}
}

@article{yan2025multiscalegraphneuralprocess,
  title   = {A Multi-Scale Graph Neural Process with Cross-Drug Co-Attention for Drug-Drug Interactions Prediction},
  author  = {Yan, Zimo and Xie, Zheng and Zhang, Jie and Song, Yiping and Li, Hao},
  journal = {arXiv preprint arXiv:2509.15256},
  year    = {2025}
}

@inproceedings{rampasek2022recipe,
  title     = {Recipe for a General, Powerful, Scalable Graph Transformer},
  author    = {Ramp{\'a}{\v{s}}ek, Ladislav and Galkin, Mikhail and Dwivedi, Vijay Prakash and Luu, Anh Tuan and Wolf, Guy and Beaini, Dominique},
  booktitle = {Advances in Neural Information Processing Systems},
  year      = {2022}
}

@article{chinta2013heat,
  title={Heat kernels on regular graphs and generalized Ihara zeta function formulas},
  author={Chinta, Gautam and Jorgenson, Jay and Karlsson, Anders},
  journal={Monatshefte f{\"u}r Mathematik},
  volume={178},
  number={2},
  pages={171--190},
  year={2015},
  publisher={Springer}
}

@article{chung1999coverings,
  title        = {Coverings, heat kernels and spanning trees},
  author       = {Chung, Fan and Yau, Shing-Tung},
  journal      = {The Electronic Journal of Combinatorics},
  volume       = {6},
  number       = {1},
  pages        = {R12},
  year         = {1999}
}

@inproceedings{sun2009hks,
  title     = {A Concise and Provably Informative Multi-Scale Signature Based on Heat Diffusion},
  author    = {Sun, Jian and Ovsjanikov, Maks and Guibas, Leonidas J.},
  booktitle = {Computer Graphics Forum},
  volume    = {28},
  number    = {5},
  pages     = {1383--1392},
  year      = {2009}
}
\clearpage
\appendix
\section{Algebraic Relation Between Distance Encoding and Laplacian Spectral Coordinates}

In this appendix, we give detailed proofs of the results stated in Section~\ref{SE5}, which connect distance encoding (DE) based on shortest-path distances and Laplacian spectral coordinates via an explicit algebraic map.

Throughout, we use the notation and standing assumptions introduced in Section~\ref{SE3} and Section~\ref{SE5}. In particular, $G = (V,E)$ is a finite, simple, connected, undirected graph, $L$ is the normalized Laplacian defined in Section~\ref{SE3}, the truncated spectral embedding $\Phi_t^{(m)}$ and the truncated diffusion distance $d_t^{(m)}$ are given there, and the random-regular model, monotone linkage, anchor general position and spectral injectivity assumptions are Assumptions~\ref{ass:Gnr_main}-\ref{ass:inj_main}.

We also use the DE-to-LapPE trilateration operator $T(\cdot)$ introduced in Definition~\ref{def:trilateration_operator}, which we briefly recall for convenience.

\subsection{Reminder of the trilateration operator and distance matrices}

Fix $t > 0$ and $m \in \mathbb{N}$, and let $a_1,\dots,a_{m+1} \in V$ be anchors with
\begin{equation}
  p_i = \Phi_t^{(m)}(a_i) \in \mathbb{R}^m, \quad i = 1,\dots,m+1.
\end{equation}
The $m \times m$ matrix $A$ and the map $b : \mathbb{R}^{m+1} \to \mathbb{R}^m$ are
\begin{equation}
  A
  =
  2
  \begin{pmatrix}
    (p_1 - p_{m+1})^\top \\
    \vdots \\
    (p_m - p_{m+1})^\top
  \end{pmatrix},
\end{equation}
\begin{equation}
  b(r)
  =
  \begin{pmatrix}
    \|p_1\|_2^2 - \|p_{m+1}\|_2^2 + r_{m+1}^{\,2} - r_1^{\,2} \\
    \vdots \\
    \|p_m\|_2^2 - \|p_{m+1}\|_2^2 + r_{m+1}^{\,2} - r_m^{\,2}
  \end{pmatrix},
\end{equation}
for any $r = (r_1,\dots,r_{m+1})^\top \in \mathbb{R}^{m+1}$.

Let $\psi : [0,R] \to \mathbb{R}_+$ be the strictly increasing function from Theorem~\ref{thm:local_monotone_link_pro}. For a vector $y = (y_1,\dots,y_{m+1})^\top \in \mathbb{R}^{m+1}$ we apply $\psi$ elementwise and write
\begin{equation}
  \bigl( \psi_*(y) \bigr)_i = \psi(y_i), \quad i = 1,\dots,m+1.
\end{equation}
For a node $v \in V$ and anchor set $S = \{a_1,\dots,a_{m+1}\}$, recall the distance encoding
\begin{equation}
  \zeta(v \mid S)
  = \bigl( \SPD(a_1,v), \dots, \SPD(a_{m+1},v) \bigr)^\top.
\end{equation}
Whenever $A$ is invertible, Definition~\ref{def:trilateration_operator} sets
\begin{equation}
  T(v)
  = A^{-1} b\bigl( \psi_*(\zeta(v \mid S)) \bigr) \in \mathbb{R}^m.
\end{equation}

In addition, the node-by-anchor shortest-path and truncated diffusion distance matrices are
\begin{equation}
  \bigl( D_{\mathrm{SPD}} \bigr)_{v,i} = \SPD(v,a_i),
\end{equation}
\begin{equation}
  \bigl( D_{\mathrm{diff}}^{(m)} \bigr)_{v,i}
  = d_t^{(m)}(v,a_i)
  = \bigl\| \Phi_t^{(m)}(v) - \Phi_t^{(m)}(a_i) \bigr\|_2,
\end{equation}
and we write $\psi_*(D_{\mathrm{SPD}})$ for the matrix obtained by applying $\psi$ elementwise.

\subsection{Proof of Theorem~\ref{thm:local_monotone_link_pro}}
\label{app:proof_local_monotone_link}

We begin by recalling a standard logarithmic-depth exploration window on random regular graphs.
This result is proved and used as a key input in the distance-encoding analysis of Li-Wang-Wang-Leskovec~\citeyearpar{li2009distance},
and we import it here as a black-box statement in order to fix a concrete $R=\Theta(\log n)$ regime in which local neighborhoods
exhibit tree-like expansion.

\begin{lemma}[Li et al.~\citeyearpar{li2009distance}]
\label{lem:log_window_imported}
Fix an integer $r\ge 3$ and let $G\sim\mathcal{G}_{n,r}$. There exists a sufficiently small constant $\epsilon>0$ such that,
with probability $1-o(n^{-3/2})$, the following holds. Choose a root $u\in V$ and define the BFS layer sets
\begin{equation}
Q_k := \{v\in V:\ \mathrm{SPD}(u,v)=k\}.
\end{equation}
Let $p_k$ denote the number of frontier half-edges incident to $Q_k$ that remain unmatched at the beginning of the $(k+1)$-th BFS step
in the configuration-model exposure. Then for every integer
\begin{equation}
k \in \Bigl(\frac{\epsilon}{5}\cdot\frac{\log n}{\log(r-1)}+1,\ \Bigl(\frac{2}{3}-\epsilon\Bigr)\cdot\frac{\log n}{\log(r-1)}\Bigr),
\end{equation}
one has
\begin{equation}
|Q_k|\ \ge\ (r-1-\epsilon)^{k-1}
\quad\text{and}\quad
p_k\ \ge\ (r-1-\epsilon)\,|Q_k|.
\end{equation}
\end{lemma}

\begin{lemma}
\label{lem:tree_radial}
Let $T_r$ be the infinite $r$-regular tree and let $t>0$. Let $L_{T_r}$ denote the (normalized) graph Laplacian on $T_r$ and
$K_t:=e^{-tL_{T_r}}$ its heat semigroup with heat kernel $k_t(x,y):=(K_t)_{xy}$. Then there exists a function
$h_t:\mathbb{N}_0\to(0,\infty)$ such that
\begin{equation}
k_t(x,y)=h_t(d(x,y)),\quad d(x,y):=\mathrm{SPD}_{T_r}(x,y).
\end{equation}
If one defines the diffusion distance on $T_r$ by
\begin{equation}
d_t(x,y)^2 := k_{2t}(x,x)+k_{2t}(y,y)-2k_{2t}(x,y),
\end{equation}
then $d_t(x,y)=\psi_{\mathrm{tree}}(d(x,y))$ with
\begin{equation}
\psi_{\mathrm{tree}}(d):=\sqrt{2\bigl(h_{2t}(0)-h_{2t}(d)\bigr)}.
\end{equation}
Moreover, for fixed $t>0$, $h_{2t}(d)$ is strictly decreasing in $d$, hence $\psi_{\mathrm{tree}}$ is strictly increasing.
\end{lemma}

\begin{proof}
Since $T_r$ is distance-transitive, for any two pairs $(x,y)$ and $(x',y')$ with $d(x,y)=d(x',y')$ there exists an automorphism
$\pi$ of $T_r$ such that $\pi(x)=x'$ and $\pi(y)=y'$. The Laplacian $L_{T_r}$ is invariant under automorphisms, hence the heat
semigroup $K_t=e^{-tL_{T_r}}$ commutes with $\pi$. Writing this invariance at the kernel level yields
$k_t(x,y)=k_t(\pi(x),\pi(y))=k_t(x',y')$, which implies the existence of a radial function $h_t$ with $k_t(x,y)=h_t(d(x,y))$.
Complete derivations and explicit formulas for $h_t$ on regular trees are given in \citeyearpar{chinta2013heat}; a covering-based treatment
relating heat kernels on the infinite regular tree and finite regular graphs is given in \citeyearpar{chung1999coverings}.

The diffusion-distance identity then follows algebraically. Since $T_r$ is vertex-transitive, $k_{2t}(x,x)=k_{2t}(y,y)=h_{2t}(0)$,
and radiality gives $k_{2t}(x,y)=h_{2t}(d(x,y))$. Substituting into the definition yields
\begin{equation}
\begin{aligned}
d_t(x,y)^2& = h_{2t}(0)+h_{2t}(0)-2h_{2t}(d(x,y))\\
&= 2\bigl(h_{2t}(0)-h_{2t}(d(x,y))\bigr),
\end{aligned}
\end{equation}
so $d_t(x,y)=\psi_{\mathrm{tree}}(d(x,y))$ with the stated $\psi_{\mathrm{tree}}$. The strict monotonicity of $h_{2t}(d)$ in $d$
for fixed $t>0$ is established in the regular-tree heat-kernel analyses cited above, which implies that $\psi_{\mathrm{tree}}$ is strictly
increasing.
\end{proof}

We now prove Theorem~\ref{thm:local_monotone_link_pro}. In addition to the geometric comparison hypothesis stated in the theorem, the only
nontrivial step is to quantify the effect of truncating the diffusion distance to its first $m$ nontrivial eigenmodes. We incorporate that
truncation calculation directly into the proof.

\begin{proof}[Proof of Theorem~\ref{thm:local_monotone_link_pro}]
Fix $t>0$ and $m\in\mathbb{N}$. Let $G=(V,E)$ be a finite connected graph with symmetric normalized Laplacian $L$.
Let $(\lambda_j,\phi_j)_{j=1}^n$ be an orthonormal eigenbasis with $0=\lambda_1\le \lambda_2\le\cdots\le\lambda_n$.
For $t>0$, define the full diffusion distance
\begin{equation}
d_t(u,v)^2 := \sum_{j=2}^{n} e^{-2t\lambda_j}\bigl(\phi_j(u)-\phi_j(v)\bigr)^2,
\end{equation}
and the truncated diffusion distance
\begin{equation}
d_t^{(m)}(u,v)^2 := \sum_{j=2}^{m+1} e^{-2t\lambda_j}\bigl(\phi_j(u)-\phi_j(v)\bigr)^2.
\end{equation}
Define the spectral tail energy
\begin{equation}
\mathrm{Tail}_t^{(m)}(u,v)
:=\Biggl(\sum_{j=m+2}^{n} e^{-2t\lambda_j}\bigl(\phi_j(u)-\phi_j(v)\bigr)^2\Biggr)^{1/2}.
\end{equation}

Fix any pair $u,v\in V$ with $\mathrm{SPD}(u,v)\le R$. We first relate $d_t^{(m)}(u,v)$ to $d_t(u,v)$ by an exact decomposition.
Splitting the defining sum of $d_t(u,v)^2$ at index $m+1$ yields
\begin{equation}
\begin{aligned}
d_t(u,v)^2
&= \sum_{j=2}^{m+1} e^{-2t\lambda_j}\bigl(\phi_j(u)-\phi_j(v)\bigr)^2 \\
 &+ \sum_{j=m+2}^{n} e^{-2t\lambda_j}\bigl(\phi_j(u)-\phi_j(v)\bigr)^2.
\end{aligned}
\end{equation}
By definition, the first sum equals $d_t^{(m)}(u,v)^2$ and the second sum equals $\mathrm{Tail}_t^{(m)}(u,v)^2$, hence
\begin{equation}
d_t(u,v)^2 = d_t^{(m)}(u,v)^2 + \mathrm{Tail}_t^{(m)}(u,v)^2.
\end{equation}
Let $a:=d_t^{(m)}(u,v)^2\ge 0$ and $b:=\mathrm{Tail}_t^{(m)}(u,v)^2\ge 0$. Then $d_t(u,v)=\sqrt{a+b}$ and $d_t^{(m)}(u,v)=\sqrt a$,
so
\begin{equation}
\begin{aligned}
&d_t(u,v)-d_t^{(m)}(u,v)=\sqrt{a+b}-\sqrt{a}
\\ &=\frac{(a+b)-a}{\sqrt{a+b}+\sqrt{a}}
=\frac{b}{\sqrt{a+b}+\sqrt{a}}.
\end{aligned}
\end{equation}
Since $\sqrt{a+b}+\sqrt{a}\ge \sqrt{a+b}\ge \sqrt{b}$, we obtain
\begin{equation}
\begin{aligned}
0 \le d_t(u,v)-d_t^{(m)}(u,v)&=\frac{b}{\sqrt{a+b}+\sqrt{a}}
\le \frac{b}{\sqrt{b}}\\&=\sqrt{b}=\mathrm{Tail}_t^{(m)}(u,v),
\end{aligned}
\end{equation}
which implies
\begin{equation}
\bigl|d_t^{(m)}(u,v)-d_t(u,v)\bigr|\le \mathrm{Tail}_t^{(m)}(u,v).
\end{equation}

We now compare $d_t^{(m)}(u,v)$ to $\psi(\mathrm{SPD}(u,v))$. By the triangle inequality,
\begin{equation}
\resizebox{\columnwidth}{!}{$
\begin{aligned}
\bigl|d_t^{(m)}(u,v)-\psi(\mathrm{SPD}(u,v))\bigr|
&\le \bigl|d_t^{(m)}(u,v)-d_t(u,v)\bigr|\\+\bigl|&d_t(u,v)-\psi(\mathrm{SPD}(u,v))\bigr|.
\end{aligned}
$}
\end{equation}
On the high-probability event from the truncation hypothesis in the theorem statement, the first term satisfies
\begin{equation}
\begin{aligned}
&\bigl|d_t^{(m)}(u,v)-d_t(u,v)\bigr|\le \mathrm{Tail}_t^{(m)}(u,v)\\
&\le
\sup_{\mathrm{SPD}(x,y)\le R}\mathrm{Tail}_t^{(m)}(x,y)\le \delta_n^{\mathrm{trunc}}.
\end{aligned}
\end{equation}
On the high-probability event from the geometric comparison hypothesis in the theorem statement, the second term satisfies
\begin{equation}
\bigl|d_t(u,v)-\psi(\mathrm{SPD}(u,v))\bigr|\le \delta_n^{\mathrm{geom}}.
\end{equation}
Intersecting the two events and combining the two inequalities yields
\begin{equation}
\bigl|d_t^{(m)}(u,v)-\psi(\mathrm{SPD}(u,v))\bigr|\le \delta_n^{\mathrm{trunc}}+\delta_n^{\mathrm{geom}}
=\delta_n^L.
\end{equation}
Because the bounds are uniform over all pairs with $\mathrm{SPD}(u,v)\le R$ on the same intersection event, the conclusion holds with
high probability over $G\sim\mathcal{G}_{n,r}$.
\end{proof}

We will use Theorem~\ref{thm:local_monotone_link_pro} in the proof of the next reconstruction theorem.

\subsection{Proof of Proposition~\ref{prop:generic_anchors}}
\label{app:proof_generic_anchors}

\begin{proof}[Proof of Proposition~\ref{prop:generic_anchors}]
Let $a_1,\dots,a_{m+1}$ be i.i.d.\ uniform on $V$ and set $p_i:=\Phi_t^{(m)}(a_i)\in\mathbb{R}^m$.
Define $M=[p_1-p_{m+1}\ \cdots\ p_m-p_{m+1}]\in\mathbb{R}^{m\times m}$.

Since $(a_1,\dots,a_{m+1})$ and $(V_1,\dots,V_{m+1})$ have the same law, the definition~\eqref{def:eta_prob} gives
\begin{equation}
\mathbb{P}\bigl(\det(M)=0\bigr)=\eta_n.
\end{equation}

Therefore $\mathbb{P}(\det(M)\neq 0)=1-\eta_n=1-o(1)$.

Assume next the jitter construction. Fix $G$ and the anchors, hence fix $p_1,\dots,p_{m+1}$. Let $\varepsilon>0$ and let
$\xi_1,\dots,\xi_{m+1}\in\mathbb{R}^m$ be i.i.d.\ with a joint density, and define $\tilde p_i:=p_i+\varepsilon\xi_i$.
Let $\tilde M=[\tilde p_1-\tilde p_{m+1}\ \cdots\ \tilde p_m-\tilde p_{m+1}]$. Then
\begin{equation}
\tilde p_i-\tilde p_{m+1}=(p_i-p_{m+1})+\varepsilon(\xi_i-\xi_{m+1}),
\end{equation}
so
\begin{equation}
\tilde M = M+\varepsilon X,
\quad
X:=[\xi_1-\xi_{m+1}\ \cdots\ \xi_m-\xi_{m+1}].
\end{equation}
Hence $\det(\tilde M)=\det(M+\varepsilon X)$ is a polynomial in the entries of $(\xi_1,\dots,\xi_{m+1})$.
This polynomial is not identically zero: indeed, choose a deterministic realization with $\xi_{m+1}=0$ and $\xi_i=e_i$ for
$i=1,\dots,m$, where $\{e_i\}$ is the standard basis of $\mathbb{R}^m$. Then $X=I_m$ and
\begin{equation}
\det(M+\varepsilon X)=\det(M+\varepsilon I_m).
\end{equation}
As a polynomial in $\varepsilon$, $\det(M+\varepsilon I_m)$ has leading term $\varepsilon^m$ and therefore is not the zero polynomial,
so there exist noise values making $\det(M+\varepsilon X)\neq 0$. Consequently the zero set
\begin{equation}
\resizebox{\columnwidth}{!}{$
Z:=\{(\xi_1,\dots,\xi_{m+1})\in(\mathbb{R}^m)^{m+1}:\ \det(M+\varepsilon X)=0\}
$}
\end{equation}
is a proper algebraic variety and has Lebesgue measure zero. Because $(\xi_1,\dots,\xi_{m+1})$ has a joint density,
\begin{equation}
\mathbb{P}\bigl((\xi_1,\dots,\xi_{m+1})\in Z\mid G,a_1,\dots,a_{m+1}\bigr)=0,
\end{equation}
which is equivalent to $\mathbb{P}(\det(\tilde M)\neq 0\mid G,a_1,\dots,a_{m+1})=1$.
This proves the affine independence under jitter.
\end{proof}

\subsection{Proof of Theorem~\ref{thm:DE_to_LapPE_pointwise}}
\label{app:proofs}
\begin{proof}[Proof of Theorem~\ref{thm:DE_to_LapPE_pointwise}]
Fix $G \sim \mathcal{G}_{n,r}$, anchors $a_1,\dots,a_{m+1}$ and a node $v \in V$ with $\SPD(a_i,v) \le R$ for all $i$. We denote
\begin{equation}
  z = \Phi_t^{(m)}(v) \in \mathbb{R}^m,
  p_i = \Phi_t^{(m)}(a_i) \in \mathbb{R}^m,
\end{equation}
and its distance encoding with respect to $S$ by
\begin{equation}
  y_i = \SPD(a_i,v),
  \quad
  y = \zeta(v \mid S)
  = (y_1,\dots,y_{m+1})^\top.
\end{equation}

\paragraph{Step 1: From shortest-path distances to approximate radii.}
Let $y_i := \SPD(a_i,v)$. By Theorem~\ref{thm:local_monotone_link_pro}, there exist a strictly
increasing function $\psi$ and a sequence $\delta_n^L=o(1)$ such that, with high probability,
for all $i$ with $y_i \le R$,
\begin{equation}
  \bigl| d_t^{(m)}(a_i,v) - \psi(y_i) \bigr|
  \le \delta_n^L.
\end{equation}
Define
\begin{equation}
  r_i := \psi(y_i),
  \quad
  r := (r_1,\dots,r_{m+1})^\top.
\end{equation}

For later use, define the exact truncated radius
\begin{equation}
  r_i^\star := d_t^{(m)}(a_i,v), \quad i = 1,\dots,m+1,
\end{equation}
and the corresponding errors
\begin{equation}
  \zeta_{i,n} := r_i^\star - r_i,
\end{equation}
so that
\begin{equation}
  r_i^\star = r_i + \zeta_{i,n},
  \quad
  |\zeta_{i,n}| \le \delta_n^L.
\end{equation}

\paragraph{Step 2: Exact spectral trilateration with $\{r_i^\star\}$}
By definition of $d_t^{(m)}$, for each anchor $a_i$ we have
\begin{equation}
  d_t^{(m)}(a_i,v)
  = \bigl\| \Phi_t^{(m)}(a_i) - \Phi_t^{(m)}(v) \bigr\|_2
  = \bigl\| p_i - z \bigr\|_2.
\end{equation}
Squaring both sides yields
\begin{equation}
  \bigl\| z - p_i \bigr\|_2^2
  = \bigl( r_i^\star \bigr)^2.
\end{equation}
Using the identity
\begin{equation}
  \bigl\| z - p_i \bigr\|_2^2
  = \|z\|_2^2 - 2 \langle z, p_i \rangle + \|p_i\|_2^2,
\end{equation}
this becomes
\begin{equation}
  \|z\|_2^2 - 2 \langle z, p_i \rangle + \|p_i\|_2^2
  = \bigl( r_i^\star \bigr)^2.
\end{equation}
These are $m+1$ equations in the unknown $z \in \mathbb{R}^m$ and the scalar $\|z\|_2^2$.

To eliminate $\|z\|_2^2$, we subtract the equation corresponding to index $m+1$ from that corresponding to a general index $i \in \{1,\dots,m\}$. For each such $i$, we consider
\begin{equation}
\resizebox{\columnwidth}{!}{$
\begin{aligned}
  &\Bigl( \|z\|_2^2 - 2 \langle z, p_i \rangle + \|p_i\|_2^2 \Bigr)
   &- \Bigl( \|z\|_2^2 - 2 \langle z, p_{m+1} \rangle + \|p_{m+1}\|_2^2 \Bigr)
  \\
  &= \bigl( r_i^\star \bigr)^2 - \bigl( r_{m+1}^\star \bigr)^2.
\end{aligned}
$}
\end{equation}
On the left-hand side, we compute term by term
\begin{equation}
\resizebox{\columnwidth}{!}{$
\begin{aligned}
  &\|z\|_2^2 - 2 \langle z, p_i \rangle + \|p_i\|_2^2
   - \|z\|_2^2 + 2 \langle z, p_{m+1} \rangle - \|p_{m+1}\|_2^2
  \\
  &= -2 \langle z, p_i \rangle + 2 \langle z, p_{m+1} \rangle
     + \|p_i\|_2^2 - \|p_{m+1}\|_2^2.
\end{aligned}
$}
\end{equation}
Factoring the inner products gives
\begin{equation}
\resizebox{\columnwidth}{!}{$
  -2 \langle z, p_i \rangle + 2 \langle z, p_{m+1} \rangle
  = 2 \langle z, p_{m+1} - p_i \rangle
  = -2 \langle z, p_i - p_{m+1} \rangle.
  $}
\end{equation}
Therefore
\begin{equation}
\resizebox{\columnwidth}{!}{$
  -2 \langle z, p_i - p_{m+1} \rangle
  + \|p_i\|_2^2 - \|p_{m+1}\|_2^2
  = \bigl( r_i^\star \bigr)^2 - \bigl( r_{m+1}^\star \bigr)^2.
  $}
\end{equation}
Multiplying both sides by $-1$ yields
\begin{equation}
\resizebox{\columnwidth}{!}{$
  2 \langle z, p_i - p_{m+1} \rangle
  = \|p_i\|_2^2 - \|p_{m+1}\|_2^2
    + \bigl( r_{m+1}^\star \bigr)^2 - \bigl( r_i^\star \bigr)^2.
    $}
\end{equation}
For each $i = 1,\dots,m$, this is a linear equation in the components of $z$.

We now stack these equations for $i = 1,\dots,m$. On the left-hand side, the $i$-th entry is
\begin{equation}
  2 (p_i - p_{m+1})^\top z.
\end{equation}
Define
\begin{equation}
  A
  =
  2
  \begin{pmatrix}
    (p_1 - p_{m+1})^\top \\
    \vdots \\
    (p_m - p_{m+1})^\top
  \end{pmatrix}.
\end{equation}
Then the stacked left-hand side is exactly $A z$.

On the right-hand side, define
\begin{equation}
  b^\star
  =
  \begin{pmatrix}
    \|p_1\|_2^2 - \|p_{m+1}\|_2^2 + (r_{m+1}^\star)^2 - (r_1^\star)^2 \\
    \vdots \\
    \|p_m\|_2^2 - \|p_{m+1}\|_2^2 + (r_{m+1}^\star)^2 - (r_m^\star)^2
  \end{pmatrix}.
\end{equation}
The $m$ linear equations can be written compactly as
\begin{equation}
  A z = b^\star.
\end{equation}

By Proposition~\ref{prop:generic_anchors}-\ref{prop:generic_anchors_jitter}, the vectors $p_1 - p_{m+1},\dots,p_m - p_{m+1}$ are linearly independent in $\mathbb{R}^m$. Therefore the rows of $A/2$ are linearly independent, hence $A$ has full rank $m$ and is invertible. There is then a unique solution
\begin{equation}
  z = A^{-1} b^\star.
\end{equation}
Since $z = \Phi_t^{(m)}(v)$, this expresses the true spectral coordinate in terms of the exact truncated radii $r_i^\star$.

\paragraph{Step 3: Relating the exact and approximate right-hand sides.}
In practice we construct $T(v)$ using $r_i$ instead of $r_i^\star$. The corresponding right-hand side is
\begin{equation}
  b(r)
  =
  \begin{pmatrix}
    \|p_1\|_2^2 - \|p_{m+1}\|_2^2 + r_{m+1}^2 - r_1^2 \\
    \vdots \\
    \|p_m\|_2^2 - \|p_{m+1}\|_2^2 + r_{m+1}^2 - r_m^2
  \end{pmatrix}.
\end{equation}
By Definition~\ref{def:trilateration_operator}, the DE-based approximation is
\begin{equation}
  T(v) = A^{-1} b(r).
\end{equation}

We now compare $b^\star$ and $b(r)$ entry by entry. For $i = 1,\dots,m$, we have
\begin{equation}
\begin{aligned}
  b^\star_i - b(r)_i
  &= \Bigl( \|p_i\|_2^2 - \|p_{m+1}\|_2^2 + (r_{m+1}^\star)^2 - (r_i^\star)^2 \Bigr)
  \\ &- \Bigl( \|p_i\|_2^2 - \|p_{m+1}\|_2^2 + r_{m+1}^2 - r_i^2 \Bigr)
  \\
  &= (r_{m+1}^\star)^2 - (r_i^\star)^2 - r_{m+1}^2 + r_i^2.
\end{aligned}
\end{equation}
Using $r_i^\star = r_i + \zeta_{i,n}$, we compute
\begin{equation}
\begin{aligned}
  (r_{m+1}^\star)^2 - r_{m+1}^2
  &= (r_{m+1} + \zeta_{m+1,n})^2 - r_{m+1}^2
  \\
  = r_{m+1}^2 &+ 2 r_{m+1} \zeta_{m+1,n} + \zeta_{m+1,n}^2 - r_{m+1}^2
  \\
  &= 2 r_{m+1} \zeta_{m+1,n} + \zeta_{m+1,n}^2,
\end{aligned}
\end{equation}
and similarly
\begin{equation}
\begin{aligned}
  (r_i^\star)^2 - r_i^2
  &= (r_i + \zeta_{i,n})^2 - r_i^2
  \\
  &= 2 r_i \zeta_{i,n} + \zeta_{i,n}^2.
\end{aligned}
\end{equation}
Substituting these expressions, we obtain
\begin{equation}
\begin{aligned}
  &(r_{m+1}^\star)^2 - (r_i^\star)^2 - r_{m+1}^2 + r_i^2\\
  = &\bigl( (r_{m+1}^\star)^2 - r_{m+1}^2 \bigr)
   - \bigl( (r_i^\star)^2 - r_i^2 \bigr)
  \\
  = &\bigl( 2 r_{m+1} \zeta_{m+1,n} + \zeta_{m+1,n}^2 \bigr)
   - \bigl( 2 r_i \zeta_{i,n} + \zeta_{i,n}^2 \bigr).
\end{aligned}
\end{equation}
Therefore
\begin{equation}
  b^\star_i - b(r)_i
  = 2 r_{m+1} \zeta_{m+1,n} + \zeta_{m+1,n}^2
    - 2 r_i \zeta_{i,n} - \zeta_{i,n}^2.
\end{equation}

We now bound $|b^\star_i - b(r)_i|$ using the inequalities
\begin{equation}
  |\zeta_{i,n}| \le \delta_n^L,
  \quad
  |\zeta_{m+1,n}| \le \delta_n^L.
\end{equation}

Because $m$ is fixed and the embedding dimension is finite, there exists a constant $B_t > 0$ such that
\begin{equation}
  \| \Phi_t^{(m)}(u) \|_2 \le B_t
\end{equation}
for all $u \in V$. Hence, for all $i$,
\begin{equation}
  r_i^\star = d_t^{(m)}(a_i,v) = \|\Phi_t^{(m)}(a_i)-\Phi_t^{(m)}(v)\|_2 \le 2B_t,
\end{equation}
and by $r_i = r_i^\star - \zeta_{i,n}$ we have
\begin{equation}
  |r_i| \le |r_i^\star| + |\zeta_{i,n}| \le 2B_t + \delta_n^L.
\end{equation}
Therefore, for all sufficiently large $n$ (so that $\delta_n^L \le 1$), there exists a constant $\widetilde{B}_t>0$
(depending only on $t,m,r$) such that
\begin{equation}
  |r_i| \le \widetilde{B}_t,
  \quad
  |r_{m+1}| \le \widetilde{B}_t
  \quad \text{for all } i=1,\dots,m.
\end{equation}

Using the triangle inequality, we obtain
\begin{equation}
\resizebox{\columnwidth}{!}{$
\begin{aligned}
  |b^\star_i - b(r)_i|
  &\le 2 |r_{m+1}| \, |\zeta_{m+1,n}| + \zeta_{m+1,n}^2
      + 2 |r_i| \, |\zeta_{i,n}| + \zeta_{i,n}^2
  \\
  &\le 2 \widetilde{B}_t \delta_n^L + (\delta_n^L)^2
      + 2 \widetilde{B}_t \delta_n^L + (\delta_n^L)^2
  \\
  &= 4 \widetilde{B}_t \delta_n^L + 2(\delta_n^L)^2.
\end{aligned}
$}
\end{equation}
For sufficiently large $n$, $\delta_n^L \le 1$ implies $(\delta_n^L)^2 \le \delta_n^L$, hence
there exists a constant $C_1>0$ (depending only on $\widetilde{B}_t$) such that
\begin{equation}
  |b^\star_i - b(r)_i| \le C_1\,\delta_n^L,\quad i=1,\dots,m.
\end{equation}

Collecting the $m$ inequalities yields
\begin{equation}
  \| b^\star - b(r) \|_2
  \le \sqrt{m}\,C_1\,\delta_n^L.
\end{equation}

\paragraph{Step 4: Stability of the linear system and conclusion.}
The true spectral coordinate and the DE-based coordinate satisfy
\begin{equation}
  A z = b^\star,
  \quad
  A T(v) = b(r).
\end{equation}
Subtracting yields
\begin{equation}
  A\bigl(z-T(v)\bigr)= b^\star-b(r).
\end{equation}
Since $A$ is invertible, we have
\begin{equation}
  z - T(v) = A^{-1}\bigl(b^\star-b(r)\bigr).
\end{equation}
Taking norms gives
\begin{equation}
  \bigl\| z - T(v) \bigr\|_2
  \le \|A^{-1}\|_{\mathrm{op}} \,\bigl\| b^\star - b(r) \bigr\|_2.
\end{equation}
Combining this with the bound on $\|b^\star-b(r)\|_2$ from Step 3 proves the claim.
Recalling that $z=\Phi_t^{(m)}(v)$ completes the proof.

\end{proof}

\subsection{Proof of Theorem~\ref{thm:DE_to_LapPE_matrix}}

\begin{proof}[Proof of Theorem~\ref{thm:DE_to_LapPE_matrix}]
We prove the Frobenius norm bound for the discrepancy between the truncated diffusion distance matrix
$D_{\mathrm{diff}}^{(m)}$ and the transformed shortest-path matrix $\psi_*(D_{\mathrm{SPD}})$.

Recall that
\begin{equation}
\begin{aligned}
 (D_{\mathrm{SPD}})_{v,i} &= \SPD(v,a_i), \\
  (D_{\mathrm{diff}}^{(m)})_{v,i} &= d_t^{(m)}(v,a_i), \\
  (\psi_*(D_{\mathrm{SPD}}))_{v,i} &= \psi(\SPD(v,a_i)).   
\end{aligned}
\end{equation}

By Theorem~\ref{thm:local_monotone_link_pro}, with high probability, for all nodes $v\in V$ and anchors $a_i$
with $\SPD(v,a_i)\le R$,
\begin{equation}
  \left|
    (D_{\mathrm{diff}}^{(m)})_{v,i} - (\psi_*(D_{\mathrm{SPD}}))_{v,i}
  \right|
  \le \delta_n^L.
\end{equation}

Therefore,
\begin{equation}
\begin{aligned}
  &\bigl\| D_{\mathrm{diff}}^{(m)} - \psi_*(D_{\mathrm{SPD}}) \bigr\|_F^2 \\
  = &\sum_{v \in V} \sum_{i=1}^{m+1}
    \left(
      (D_{\mathrm{diff}}^{(m)})_{v,i}
     - (\psi_*(D_{\mathrm{SPD}}))_{v,i}
    \right)^2
  \\
  \le& \sum_{v \in V} \sum_{i=1}^{m+1} (\delta_n^L)^2 = n(m+1)(\delta_n^L)^2.
\end{aligned}
\end{equation}
Taking square roots yields
\begin{equation}
  \bigl\| D_{\mathrm{diff}}^{(m)} - \psi_*(D_{\mathrm{SPD}}) \bigr\|_F
  \le \delta_n^L \sqrt{n(m+1)}.
\end{equation}
This proves Theorem~\ref{thm:DE_to_LapPE_matrix}.
\end{proof}

\section{Experimental Details}
\label{app:exp-details}

\subsection{Datasets}
\label{app:exp-details-datasets}

We evaluate all methods on two widely used benchmark datasets for drug-drug interaction prediction: DrugBank and ChCh-Miner. In both cases, we follow the inductive link prediction protocol adopted in previous DDI work and in the MPNP-DDI framework \citeyearpar{yan2025multiscalegraphneuralprocess}.

\paragraph{DrugBank.}
The DrugBank interaction network is derived from the DrugBank 5.\,0 database and has been widely adopted in recent DDI studies \citeyearpar{ryu2018deep,niu2024dasddi,wang2024structnetddi}. The graph nodes correspond to small-molecule drugs and edges indicate documented interactions. Following standard preprocessing pipelines \citeyearpar{niu2024dasddi,wang2024structnetddi}, we obtain a graph with approximately $1{,}700$ drugs and $190{,}000$ labeled drug pairs, grouped into $86$ interaction types in the original resource. In this work, we focus on the binary link prediction setting and treat each pair as either interacting or non-interacting. For each drug, we construct a molecular graph from its SMILES string using RDKit, where atoms are nodes with categorical and numerical attributes (such as atom type, degree, formal charge, aromaticity), and chemical bonds are edges with bond-type features.

\paragraph{ChCh-Miner.}
The ChCh-Miner dataset is a medium-scale DDI network curated from approved drugs and released as part of the BioSNAP collection \citeyearpar{zitnik2018modeling}. Nodes represent drugs and edges indicate known interactions. The network contains $1{,}514$ drugs and $48{,}514$ documented DDI links \citeyearpar{niu2024dasddi,wang2024structnetddi}. As with DrugBank, we treat DDI prediction as a binary link prediction problem on this drug-drug graph. Molecular graphs for individual drugs are constructed from SMILES in the same way as for DrugBank.

\paragraph{Data splits.}
We follow the splitting strategy in MPNP-DDI \citeyearpar{yan2025multiscalegraphneuralprocess}. All known interactions are randomly split into training, validation, and test sets at the level of drug pairs. Negative examples are generated by uniformly sampling unobserved drug pairs, maintaining a fixed positive-to-negative ratio in each split, as is standard in DDI link prediction \citeyearpar{niu2024dasddi,he2022mffgnn}. We adopt an inductive setting where a subset of drugs appears only in the validation or test sets, so that models must generalize to unseen molecules rather than memorizing specific pairs.

\subsection{Baselines and Model Variants}
\label{app:exp-details-model}

Because our focus is on understanding the role of positional encodings within a fixed architecture, we keep the backbone model identical across settings and vary only the positional encoding module. Concretely, we consider three primary variants (NoPE/DE/LapPE) for controlled comparisons, and additionally report two widely used PE baselines (RWSE and HKS) as sanity-check references under the same backbone and training protocol.

\begin{itemize}
    \item \textbf{NoPE.} The original MPNP-DDI model \citeyearpar{yan2025multiscalegraphneuralprocess} without any explicit positional encodings. Node features are given solely by atom-level descriptors derived from RDKit, and the model relies on its multi-scale message-passing scheme to infer structural information.

    \item \textbf{DE.} The backbone model augmented with distance encodings. For each molecular graph, we sample $k$ anchor atoms and compute shortest-path distances from every node to these anchors. A radial transformation $\psi(\cdot)$ is applied to each distance, and the resulting vectors are concatenated to the original node features. The DE module is shared across all experiments, and its design is studied in detail in the ablation experiments.

    \item \textbf{LapPE.} The backbone model augmented with Laplacian positional encodings. For each molecular graph, we compute the first $m$ non-trivial eigenvectors of the normalized Laplacian and concatenate them to the node features, following standard practice in spectral GNNs \citeyearpar{dwivedi2023benchmarking}. We choose $m$ to match the dimensionality of the DE features so that all variants have comparable parameter counts.

    \item \textbf{RWSE (reference baseline).} The backbone model augmented with random-walk structural encodings (RWSE), where each node is assigned a vector of $K$-step random-walk return probabilities and the resulting encoding is concatenated to node features, following common practice in graph Transformers and PE benchmarks\citeyearpar{rampasek2022recipe}. We use steps $\mathcal{T}=\{1,2,4,8,16\}$.

    \item \textbf{HKS (reference baseline).} The backbone model augmented with heat-kernel signatures (HKS)\citeyearpar{sun2009hks}, constructed from a truncated eigenspace of the normalized Laplacian. Specifically, for each node we compute a diffusion-time embedding with times $\mathcal{S}=\{0.1,0.5,1,2,5\}$ using the top-$m$ eigenpairs and concatenate it to node features. We use a truncated eigenspace dimension of $m{=}32$.
\end{itemize}

For fair comparison, all PE variants are used as plug-in augmentations to node features and trained with the same backbone, optimizer, and training budget. When applicable, we choose PE dimensionalities to be of comparable scale to the DE/LapPE feature augmentation.

All models are implemented in PyTorch \citeyearpar{paszke2019pytorch} using PyTorch Geometric for efficient graph operations \citeyearpar{fey2019fast}. Unless otherwise specified, we follow the main architectural and optimization choices of MPNP-DDI \citeyearpar{yan2025multiscalegraphneuralprocess}.

\subsubsection{Feature Normalization.}

For molecular graphs, we construct standard atom and bond features using RDKit, including atom type, degree, valence, aromaticity, hybridization, and bond type. Categorical attributes are encoded as one-hot vectors and concatenated with numerical descriptors. We apply feature-wise standardization (zero mean, unit variance) across the training set for all continuous features and reuse the same statistics at validation and test time.

For Laplacian positional encodings, raw eigenvectors are normalized within each molecular graph to have zero mean and unit variance per eigenvector. To reduce sensitivity to global sign flips, we optionally add small Gaussian noise during training. For distance encodings, shortest-path distances are first rescaled by the median non-zero distance within each graph and then passed through a radial map $\psi(\cdot)$. We consider three choices in our experiments: $\psi(d) = d$, $\psi(d) = \exp(-d)$, and $\psi(d) = \log(1 + d)$. The transformed distances are further standardized across the training set.

\subsubsection{Architecture Design.}

The backbone architecture follows the multi-scale Graph Neural Process design of MPNP-DDI \citeyearpar{yan2025multiscalegraphneuralprocess}. Each drug is represented by both its original molecular graph and its corresponding line graph, which captures bond-level interactions. Node and edge features (including positional encodings when present) are projected to a hidden dimension of 64. The model stacks three Graph Neural Process blocks; each block runs two iterations of message passing on both the molecular graph and the line graph, followed by aggregation and cross-scale fusion, yielding a hierarchy of representations from local substructures to global topology.

For a pair of drugs, the block-wise representations are fed into a cross-drug co-attention module, which computes context-aware embeddings for each drug conditioned on its partner. These pairwise embeddings are then passed through a multilayer perceptron to predict the probability of an interaction. DE and LapPE variants differ only in the additional node-level inputs; all subsequent processing, including the co-attention and decoder, is shared.

\subsubsection{Optimization Settings.}

We train all models with the AdamW optimizer \citeyearpar{loshchilov2019decoupled} using a cosine annealing learning rate scheduler over 50 epochs, following \citeyearpar{yan2025multiscalegraphneuralprocess}. The initial learning rate and weight decay are selected by grid search on the validation set within a standard range (for example, learning rates between $10^{-4}$ and $10^{-3}$ and weight decay between $10^{-5}$ and $10^{-3}$). We use binary cross-entropy loss on the predicted interaction probabilities.

Early stopping is applied based on validation AUROC: if no improvement is observed for 10 consecutive epochs, training is terminated and the best checkpoint is kept. All experiments are repeated with three random seeds, and we report the mean and standard deviation across runs.

\subsubsection{Batch Size and Gradient Variance.}

Due to the large number of drug pairs and the memory cost of the multi-scale backbone, we use a mini-batch size of 32 drug pairs and accumulate gradients over 4 steps, resulting in an effective batch size of 128, as in \citeyearpar{yan2025multiscalegraphneuralprocess}. This reduces the variance of stochastic gradients without exceeding GPU memory constraints. We also apply gradient clipping at a fixed maximum norm to prevent rare exploding gradients caused by highly connected molecular graphs.

\subsection{Evaluation Measures}
\label{app:exp-details-eval}

We treat DDI prediction as a binary classification problem on drug pairs. For both DrugBank and ChCh-Miner, we evaluate models using the Area Under the Receiver Operating Characteristic Curve (AUROC) and the F1 score, which are standard metrics in DDI prediction \citeyearpar{niu2024dasddi,he2022mffgnn,wang2024structnetddi}. AUROC captures the ranking quality over all thresholds, while F1 summarizes the trade-off between precision and recall at a specific threshold.

During training and model selection, we monitor AUROC on the validation set. For F1, we select a threshold that maximizes validation F1 and apply the same threshold to the held-out test set. All reported numbers are computed on the test split using the checkpoint with the best validation AUROC and are averaged over three independent runs.

\subsection{Details of the Ablation Study}
\label{app:exp-details-ablation}

We conduct a focused ablation study to isolate the impact of key design choices in the distance encoding module. Specifically, we examine (i) the choice of radial transformation $\psi(\cdot)$ applied to shortest-path distances on DrugBank while fixing the number of anchors, and (ii) the number of anchors $k$ on ChCh-Miner while fixing $\psi(\cdot)$.

On DrugBank, we fix the number of anchors and compare three radial functions: $\psi(d) = d$, $\psi(d) = \exp(-d)$, and $\psi(d) = \log(1 + d)$. On ChCh-Miner, we fix $\psi(d) = \exp(-d)$ and vary $k \in \{4, 8, 16, 32\}$ to study the trade-off between approximation quality and model complexity. All ablation runs use the same data preprocessing, model architecture, optimizer, batch size, and early stopping criteria as the main experiments. Detailed numerical results and further analysis are reported in Section~\ref{subsec:de-ablation-results}. For additional implementation details (datasets, baselines, and hyperparameters), please refer to \url{https://anonymous.4open.science/r/Bridging-Distance-and-Spectral-Positional-Encodings-E48B}.

\end{document}